\documentclass{article}
\usepackage{amssymb,hyperref}
\usepackage[english]{babel}
\usepackage{amsmath, graphics}
\usepackage{geometry}
\usepackage{color}
\newcommand{\mathsym}[1]{{}}
\newcommand{\unicode}[1]{{}}

\usepackage{enumitem}
\setlist[enumerate,itemize]{leftmargin=4em,itemindent=0em}

\def\noi{\noindent}

\def\nqq{\hspace*{-2em}}

\def\lal{&&\nqq {}}

\def\barr{\left(\begin{array}}
\def\earr{\end{array}\right)}
\def\beq#1{\begin{equation}\label{#1}}
\def\eeq{\end{equation}}
\def\ber#1{\begin{eqnarray}\label{#1} &&\nqq}
\def\eer{\end{eqnarray}}

\newcommand{\bear}[1]{\begin{eqnarray}\label{#1}}
\newcommand{\bearr}[1]{\begin{eqnarray}\lal \label{#1}}
\newcommand{\ear}{\end{eqnarray}}

\newcommand{\R}{ {\mathbb R} }

\newcommand{\fnm}{\footnotemark}
\newcommand{\fnt}{\footnotetext}

\begin{document}

 \vspace{15pt}

 \begin{center}
 \large\bf

On fluxbrane polynomials for generalized Melvin-like solutions associated with rank 5 Lie algebras

 \vspace{15pt}

 \normalsize\bf
      S. V. Bolokhov\fnm[1]\fnt[1]{bolokhov-sv@rudn.ru}$^{,a}$
      and  V. D. Ivashchuk\fnm[2]\fnt[2]{ivas@vniims.ru}$^{,a,b}$

 \vspace{7pt}

 \it 
 
 (a)  \  Institute of Gravitation and Cosmology, \\
   \ Peoples' Friendship University of Russia 
   (RUDN University), \\ 
   6 Miklukho-Maklaya Street,
   Moscow, 117198, Russian Federation \\
 (b) \ \ \ Center for Gravitation and Fundamental Metrology, \\
  All-Russian Research Institute of Metrological Service (VNIIMS), \\ 
  46 Ozyornaya St., Moscow 119361, Russian Federation \\

 \end{center}
 \vspace{15pt}

 \small\noi

 \begin{abstract}
 We consider generalized Melvin-like solutions corresponding to Lie algebras of rank $5$ ($A_5$, $B_5$, $C_5$, $D_5$). The  solutions take place in $D$-dimensional gravitational model with five Abelian 2-forms and five scalar fields. They are governed by five moduli functions $H_s(z)$ ($s = 1,...,5$) of squared radial coordinate $z=\rho^2$ which obey five differential master equations. The moduli functions are polynomials of powers $(n_1, n_2, n_3, n_4, n_5) = (5,8,9,8,5), (10,18,24,28,15), (9,16,21,24,25), (8,14,18,10,10)$ for Lie algebras $A_5$, $B_5$, $C_5$, $D_5$ respectively. The asymptotic behaviour for the polynomials at large distances is governed by some integer-valued $5 \times 5$ matrix $\nu$ connected in a certain way with the inverse Cartan matrix of the Lie algebra and (in $A_5$ and $D_5$ cases) with the matrix representing a generator of the $\mathbb{Z}_2$-group of symmetry of the Dynkin diagram. The symmetry  and duality identities for polynomials are obtained, as well as asymptotic relations for solutions at large distances. 
  
  \end{abstract}


Math. Class. Codes: 11C08; 17B80; 17B81; 34A05; 35Q75; 70S99. 

\large 

 \section{Introduction}

  In this article, we deal with higher dimensional generalization of Melvin's solution \cite{Melv}, 
  which was studied earlier in ref. \cite{GI-09}. 
    
  The model from ref. \cite{GI-09}
  is described by metric, $n$ Abelian 2-forms and  $l \geq n$ scalar fields. 
  Here we study special solutions with $n =l =5$, which are governed by a $5 \times 5$ Cartan matrices  
  $(A_{i j})$ corresponding to Lie algebras of rank $5$: $A_5$, $B_5$,  $C_5$, $D_5$. 
  We note that ref. \cite{GI-09} contain a  special subclass of 
  fluxbrane solutions from ref. \cite{Iflux}. 
  
  We note that  Melvin's  solution in $4$-dimensional space-time describes the gravitational field of a magnetic flux tube. The multidimensional analogue of such a flux tube, supported by a certain configuration of   fields of forms, is  referred to as a fluxbrane. The appearance of fluxbrane solutions was motivated in past decades by superstring/M-theory models. A physical relevance of such solutions is that they supply an appropriate background geometry for studying various processes   which involve branes, instantons, Kaluza--Klein monopoles, pair production of magnetically charged  black holes and other configurations which can be studied with a special kind of Kaluza--Klein reduction  of a certain higher dimensional model in the presence of $U(1)$ isometry subgroup.
  (The readers who are interested in    generalizations of the Melvin solution and  fluxbrane solutions may be addressed
  to refs.  \cite{BronShikin}-\cite{Ivas-Symmetry-17}  and references therein.)  

  The fluxbrane solutions from ref. \cite{Iflux}  were  described by moduli functions $H_s(z) > 0$ 
  defined on  $(0, +\infty)$, where $z = \rho^2$ and $\rho$ is a proper radial coordinate. The moduli functions $H_s(z)$ 
  were obeying $n$  master equations (equivalent to Toda-like equations) governed by a matrix $(A_{s s'})$, 
  and the following boundary conditions were imposed: $H_{s}(+ 0) = 1$,   $s = 1,...,n$.  

  In  ref. \cite{GI-09} the matrix $(A_{s s'})$ was assumed to be coinciding with a Cartan matrix for some 
 simple finite-dimensional Lie algebra $\cal G$ of rank $n$.  In this case 
 according to  conjecture from Ref. \cite{Iflux} 
   the solutions to master equations with the boundary conditions $H_{s}(+ 0) = 1$ imposed 
   are  polynomials 
  \beq{1.3}
   H_{s}(z) = 1 + \sum_{k = 1}^{n_s} P_s^{(k)} z^k.
  \eeq
  Here $P_s^{(k)}$ are constants, $P_s^{(n_s)} \neq 0$  and 
 \beq{1.4}
  n_s = 2 \sum_{s' =1}^{n} A^{s s'}.
 \eeq 
 with the notation assumed: $(A^{s s'}) = (A_{s s'})^{-1}$.
 Here  $n_s$ are integer numbers which are components of the twice dual
 Weyl vector in the basis of simple coroots \cite{FS}.
 
The functions $H_s$ (so-called ``fluxbrane polynomials'')  describe a special solution to open Toda chain equations \cite{K,OP} which correspond to  simple finite-dimensional Lie algebra $\cal G$ \cite{I-14}.

Here we study  the solutions corresponding to Lie algebras of rank $5$.
We prove some symmetry properties, as well as the so-called duality relations of fluxbrane polynomials. The duality relations describe a behaviour of the solutions under the inversion  $\rho \to 1/\rho$. They can be mathematically understood in terms of the groups of symmetry  of Dynkin diagrams for the corresponding Lie algebras. 
For this work these groups of symmetry are   either identical ones (for Lie algebras $B_5$, $C_5$) or isomorphic to the group $\mathbb{Z}_2$    (for Lie algebras $A_5$, $D_5$).  
   The duality identities may be used in deriving a $1/\rho$-expansion for solutions at large distances $\rho$.   The corresponding asymptotic behavior of the solutions is presented. 
   
   The analogous consideration was performed earlier for the case of  Lie algebras of rank $2$: 
   $A_2$, $B_2 = C_2$, $G_2$  in ref. \cite{BolIvas-R2-17}, and for  Lie algebras of rank $3$: $A_3$, $B_3$, $C_3$ in ref. \cite{BolIvas-R3-18}, for rank-4 non-exceptional Lie algebras $A_4$, $B_4$, $C_4$, $D_4$ 
   in refs. \cite{Bol-Ivas-R4-18,Bol-Ivas-R4-19} and for exceptional Lie algebra $F_4$ in \cite{Bol-Ivas-R4-19}. Also, in ref. \cite{BolIvas-17}  the conjecture from ref. \cite{Iflux} was verified for the Lie algebra $E_6$ and certain duality relations for six $E_6$-polynomials were found.

\section{The set up and generalized Melvin solutions}

We deal with the  (smooth) manifold
\beq{2.2}
  M = (0, + \infty)  \times M_1 \times M_2,
 \eeq
 where $M_1 = S^1$ and $M_2$ is a $(D-2)$-dimensional manifold of signature $(-,+,...,+)$ which is 
 supposed to be Ricci-flat.

 The action of the model reads 
 \beq{2.1}
 S=\int d^Dx \sqrt{|g|} \biggl \{R[g]-
 \delta_{a b} g^{MN}\partial_M \varphi^{a} \partial_N \varphi^{b} - \frac{1}{2}
 \sum_{s =1}^{5}\exp[2 \vec{\lambda}_s \vec{\varphi}](F^s)^2 \biggr \}.
 \eeq
 Here $g=g_{MN}(x)dx^M\otimes dx^N$ is a (smooth) metric defiuned on $M$,
 $\vec{\varphi} = (\varphi^a)\in \R^5$ is vector which consists of scalar fields,
   $ F^s =    dA^s
          =  \frac{1}{2} F^s_{M N}  dx^{M} \wedge  dx^{N}$
 is a form of rank $2$,  $\vec{\lambda}_s = (\lambda_{s}^{a}) \in \R^5$ is vector 
 of dilatonic  coupling constants,
   $s = 1,...,5$; $a =1,...,5$.
 In (\ref{2.1}) we denote $|g| \equiv  |\det (g_{MN})|$, 
 $(F^s)^2 \equiv F^s_{M_1 M_{2}} F^s_{N_1 N_{2}}  g^{M_1 N_1} g^{M_{2} N_{2}}$.

We study a family of exact
 solutions to the field equations which correspond to the action
(\ref{2.1}) and depend on the radial coordinate $\rho$. These  
  solutions read as follows \cite{GI-09} (for more general
  fluxbrane solutions see \cite{Iflux})
 \bear{2.30}
  g= \Bigl(\prod_{s = 1}^{5} H_s^{2 h_s /(D-2)} \Bigr)
  \biggl\{  d\rho \otimes d \rho  +
  \Bigl(\prod_{s = 1}^{5} H_s^{-2 h_s} \Bigr) \rho^2 d\phi \otimes d\phi +
    g^2  \biggr\},
 \\  \label{2.31}
  \exp(\varphi^a)=
  \prod_{s = 1}^{5} H_s^{h_s  \lambda_{s}^a},
 \\  \label{2.32a}
  F^s =  q_s \left( \prod_{l = 1}^{5}  H_{l}^{- A_{s
  l}} \right) \rho d\rho \wedge d \phi,
  \ear
  $s, a = 1, ... , 5$, where  $g^1 = d\phi \otimes d\phi$ is a
  metric on one-dimensional circle $M_1 = S^1$ and $g^2$ is a  metric of 
  signatute $(-,+, \dots, +)$ on the manifold  $M_{2}$ which is supposed to be Ricci-flat.
   Here $q_s \neq 0$ are constants.  
 
 In what follows we denote $z = \rho^2$. 
 Here the functions $H_s(z) > 0$ obey the set of non-linear  equations \cite{GI-09}
\beq{1.1}
  \frac{d}{dz} \left( \frac{ z}{H_s} \frac{d}{dz} H_s \right) =
   P_s \prod_{l = 1}^{5}  H_{l}^{- A_{s l}},
  \eeq
 with  the boundary conditions imposed
 \beq{1.2}
   H_{s}(+ 0) = 1,
 \eeq
 where
 \beq{2.21}
  P_s =  \frac{1}{4} K_s q_s^2,
 \eeq
 $s = 1,..., 5$.  The condition  (\ref{1.2}) prevents a possible appearance of the conic singularity 
  for the metric at $\rho =  +0$.
 
 The parameters $h_s$  obey the following relations
  \beq{2.16}
  h_s = K_s^{-1}, \qquad  K_s = B_{s s} > 0,
  \eeq
 where
 \beq{2.17}
  B_{s l} \equiv
  1 +\frac{1}{2-D}+  \vec{\lambda}_{s} \vec{\lambda}_{l} ,
  \eeq
 $s, l = 1,..., 5$.
 The formulae for the solutions contain the 
   so-called ``quasi-Cartan'' matrix
 \beq{2.18}
  (A_{s l}) = \left( 2 B_{s l}/B_{l l} \right).
 \eeq
 

Here we study a multidimensional generalization of Melvin's solution \cite{Melv} for the case of five scalar fields and five $2$-forms. In the case when  scalar fields are absent the original Melvin's solution 
may be obtained here for  $D = 4$, one (electromagnetic) $2$-form, $M_1 = S^1$ ($0 < \phi <  2 \pi$),  $M_2 = \R^2$ and $g^2 = -  dt \otimes dt + d x \otimes d x$.

\section{Solutions related to  simple classical rank-5 Lie algebras}

Here we deal with solutions corresponding to  Lie algebras ${\cal G}$ of rank $5$. 
In this case the matrix  $A = (A_{sl})$ should coincide with one of the Cartan matrices 
\begin{gather} 
     \left(A_{ss'}\right) =
     \begin{pmatrix}
     2 & -1 & 0 & 0 & 0 \\
      -1 & 2 & -1 & 0 & 0 \\
      0 & -1 & 2 & -1 & 0 \\
      0 & 0 & -1 & 2 & -1 \\
      0 & 0 & 0 & -1 & 2 \\
     \end{pmatrix},\!
  \quad
   \begin{pmatrix}
    2 & -1 & 0 & 0 & 0 \\
     -1 & 2 & -1 & 0 & 0 \\
     0 & -1 & 2 & -1 & 0 \\
     0 & 0 & -1 & 2 & -2 \\
     0 & 0 & 0 & -1 & 2 \\
  \end{pmatrix}, 
  \quad \nonumber \\
  \begin{pmatrix}
      2 & -1 & 0 & 0 & 0 \\
      -1 & 2 & -1 & 0 & 0 \\
      0 & -1 & 2 & -1 & 0 \\
      0 & 0 & -1 & 2 & -1 \\
      0 & 0 & 0 & -2 & 2 \\
    \end{pmatrix}, 
   \quad 
  \begin{pmatrix}
        2 & -1 & 0 & 0 & 0 \\
         -1 & 2 & -1 & 0 & 0 \\
         0 & -1 & 2 & -1 & -1 \\
         0 & 0 & -1 & 2 & 0 \\
         0 & 0 & -1 & 0 & 2 \\
    \end{pmatrix}.
  \label{A.5}
\end{gather}
for ${\cal G} = A_5, \, B_5, \, C_5, \, D_5$, respectively.

The graphical presentation of these matrices by Dynkin diagrams is given in Fig. 1 .
 \vspace{25pt}
  
  \setlength{\unitlength}{1mm}
 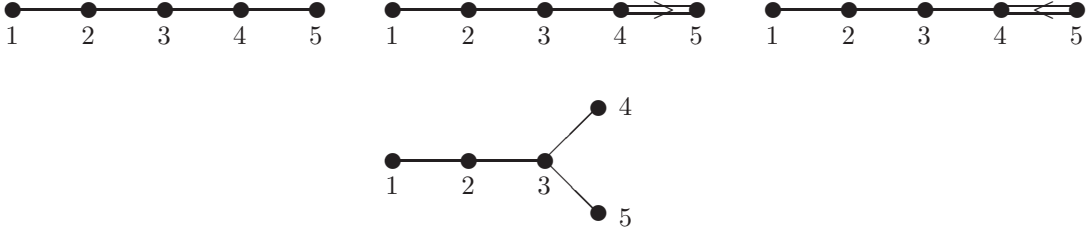
\begin{figure}[h]
 \centering
 \begin{picture}(140, 25)
 \put(2,25){\circle*{2}}
 \put(12,25){\circle*{2}}
 \put(22,25){\circle*{2}}
 \put(32,25){\circle*{2}}
 \put(42,25){\circle*{2}}
 \put(1,25){\line(1,0){40}}
  \put(0.3,20.5){$1$}
 \put(10.3,20.5){$2$}
 \put(20.3,20.5){$3$}
 \put(30.3,20.5){$4$}
 \put(40.3,20.5){$5$}
 \put(52,25){\circle*{2}}
 \put(62,25){\circle*{2}}
 \put(72,25){\circle*{2}}
 \put(82,25){\circle*{2}}
 \put(92,25){\circle*{2}}
 \put(52,25){\line(1,0){30}}
 \put(82,25.5){\line(1,0){10}}
 \put(82,24.5){\line(1,0){10}}
 \put(50.3,20.5){$1$}
 \put(60.3,20.5){$2$}
 \put(70.3,20.5){$3$}
 \put(80.3,20.5){$4$}
 \put(90.3,20.5){$5$}
 \put(85.3,23.9){\large $>$}
 \put(102,25){\circle*{2}}
 \put(112,25){\circle*{2}}
 \put(122,25){\circle*{2}}
 \put(132,25){\circle*{2}}
 \put(142,25){\circle*{2}}
 \put(102,25){\line(1,0){30}}
 \put(132,25.5){\line(1,0){10}}
 \put(132,24.5){\line(1,0){10}}
 \put(100.3,20.5){$1$}
 \put(110.3,20.5){$2$}
 \put(120.3,20.5){$3$}
 \put(130.3,20.5){$4$}
 \put(140.3,20.5){$5$}
 \put(135.3,23.9){\large $<$}
 \put(52,05){\circle*{2}}
 \put(62,05){\circle*{2}}
 \put(72,05){\circle*{2}}
 \put(79,12){\circle*{2}}
 \put(79,-2){\circle*{2}}
 \put(51,05){\line(1,0){20}}
 \put(71.3,04.5){\line(1,1){7.5}}
 \put(72,05){\line(1,-1){7.5}}
 \put(50.3,0.5){$1$}
 \put(60.3,0.5){$2$}
 \put(70.3,0.5){$3$}
 \put(81,11){$4$}
 \put(81,-3.7){$5$}
\end{picture}
  \vspace{10pt}
  \caption{Dynkin diagrams for the Lie algebras $A_5, \, B_5, \, C_5, \, D_5$, respectively.}
 \end{figure}
     
  Due to (\ref{2.16})-(\ref{2.18}) we obtain
    \begin{equation}
           \label{3.17}
    K_s =   \frac{D - 3}{D -2} +  \vec{\lambda}_{s}^2,
      \end{equation} 
    where $h_s =  K_s^{-1}$, and  
      \beq{3.18}
        \vec{\lambda}_{s} \vec{\lambda}_{l} = 
            \frac{1}{2} K_l A_{sl}  - \frac{D - 3}{D -2} \equiv G_{sl},
      \eeq    
     $s,l = 1, 2, 3, 4$;  (\ref{3.17}) is a special case of  (\ref{3.18}).

     
     {\bf Polynomials.}
    Due to the conjecture from Ref. \cite{Iflux}, the set of moduli functions  $H_1(z), ..., H_5(z)$
    which obey Eqs. (\ref{1.1}) and (\ref{1.2})
         with any matrix $A =  (A_{sl})$ from (\ref{A.5}), 
         are polynomials. Due to relation (\ref{1.4}) the powers of these polynomials 
         are following ones: 
         $(n_1, n_2, n_3, n_4, n_5) = (5,8,9,8,5), (10,18,24,28,15), 
         (9,16,21,24,25), (8,14,18,10,10)$  
         for Lie algebras $A_5$, $B_5$, $C_5$, $D_5$ respectively.
         
         Here we verify (i. e. prove) the polynomial conjecture from Ref. \cite{Iflux} just
         by solving the set of algebraic equations for the coefficients of the polynomials (\ref{1.3}) 
         which follow from master equations (\ref{1.1}). 
         
         In what follows (in this section) we present structures (or ``truncated versions'')  of these polynomials. In Appendix we present the total list of these polynomials which were obtained by using a certain MATHEMATICA algorithm. 
         Given Cartan matrix $A_{sl}$, this algorithm uses a polynomial ansatz \eqref{1.3} for $H_s(z)$ to write and solve the equations \eqref{1.1} as a system of non-linear algebraic equations on the corresponding polynomial coefficients $P_s^{(k)}$. The problem is that in case of higher ranks this system becomes quite complicated so that direct use of built-in '\textit{solve}'-like commands causes computational fails. To be more efficient, the algorithm uses the adapted computational procedure based on a certain recurrence property of the algebraic system under consideration. According to this property, among the full set of variables $P_s^{(k)}$ ($k=1,...,n_s$) one can single out a certain ``starting'' subset $P_s^{(k_0)}$ obeying the closed subsystem of equations, and resolve the remaining equations on each $k$-th step using the variables found on the previous step. As soon as all variables are found, the algorithm writes down the obtained fluxbrane polynomials and checks the correctness of the obtained solution by direct substitution into the original equations. After that, the algorithm directly verifies symmetry and duality properties for the obtained polynomials, which are discussed below.
         
         Here as in Ref. \cite{I-14}  we use the rescaled parameters
         \beq{3.P}
            p_s = P_s/n_s.    
         \eeq

\noindent{\bf $A_5$-case.} 
For the Lie algebra $A_5 $ the polynomials have the following structures
 \begin{itemize} 
\item[$H_1=$]\(
1+5 p_1 \textcolor{red}{z}+10 p_1 p_2 \textcolor{red}{{z}^2}+10 p_1 p_2 p_3 \textcolor{red}{{z}^3}+5 p_1 p_2 p_3 p_4 \textcolor{red}{{z}^4}+p_1 p_2 p_3 p_4 p_5 \textcolor{red}{{z}^5},
\)
\item[$H_2=$]\(
1+8 p_2 \textcolor{red}{z}+(10 p_1 p_2 + 18 p_2 p_3) \textcolor{red}{{z}^2}+
\dots
+(10 p_1 p_2^2 p_3^2 p_4+18 p_1 p_2^2 p_3 p_4 p_5) \textcolor{red}{{z}^6}+8 p_1 p_2^2 p_3^2 p_4 p_5 \textcolor{red}{{z}^7}+p_1 p_2^2 p_3^2 p_4^2 p_5 \textcolor{red}{{z}^8}\),
\item[$H_3=$]\(
1+9 p_3 \textcolor{red}{z}+(18 p_2 p_3+18 p_3 p_4) \textcolor{red}{{z}^2}+
\dots
+(18 p_1 p_2^2 p_3^2 p_4 p_5+18 p_1 p_2 p_3^2 p_4^2 p_5) \textcolor{red}{{z}^7}
+9 p_1 p_2^2 p_3^2 p_4^2 p_5 \textcolor{red}{{z}^8}+p_1 p_2^2 p_3^3 p_4^2 p_5 \textcolor{red}{{z}^9}\),
\item[$H_4=$]\(
1+8 p_4 \textcolor{red}{z}+(18 p_3 p_4+10 p_4 p_5) \textcolor{red}{{z}^2}+ \dots 
+(18 p_1 p_2 p_3 p_4^2 p_5+10 p_2 p_3^2 p_4^2 p_5) \textcolor{red}{{z}^6}+8 p_1 p_2 p_3^2 p_4^2 p_5 \textcolor{red}{{z}^7}+p_1 p_2^2 p_3^2 p_4^2 p_5 \textcolor{red}{{z}^8}\),
\item[$H_5=$]\(
1+5 p_5 \textcolor{red}{z}+10 p_4 p_5 \textcolor{red}{{z}^2}+10 p_3 p_4 p_5 \textcolor{red}{{z}^3}+5 p_2 p_3 p_4 p_5 \textcolor{red}{{z}^4}+p_1 p_2 p_3 p_4 p_5 \textcolor{red}{{z}^5}.
\)
\end{itemize}

\bigskip
\noindent {\bf $B_5$-case.}
For the Lie algebra $B_5 $ we obtain the following structures of  polynomials
\begin{itemize}
\item[$H_1=$]\(
1+10 p_1 \textcolor{red}{z}+45 p_1 p_2 \textcolor{red}{{z}^2}+ \dots
+45 p_1 p_2 p_3^2 p_4^2 p_5^2 \textcolor{red}{{z}^8}+10 p_1 p_2^2 p_3^2 p_4^2 p_5^2 \textcolor{red}{{z}^9}+p_1^2 p_2^2 p_3^2 p_4^2 p_5^2 \textcolor{red}{{z}^{10}},
\)
\item[$H_2=$]\(
1+18 p_2 \textcolor{red}{z}+(45 p_1 p_2+108 p_2 p_3) \textcolor{red}{{z}^2}+ \dots
+(108 p_1^2 p_2^3 p_3^3 p_4^4 p_5^4+45 p_1 p_2^3 p_3^4 p_4^4 p_5^4) \textcolor{red}{{z}^{16}}
+18 p_1^2 p_2^3 p_3^4 p_4^4 p_5^4 \textcolor{red}{{z}^{17}}+p_1^2 p_2^4 p_3^4 p_4^4 p_5^4 \textcolor{red}{{z}^{18}},
\)
\item[$H_3=$]\(
1+24 p_3 \textcolor{red}{z}+(108 p_2 p_3+168 p_3 p_4) \textcolor{red}{{z}^2}+
\dots
+(168 p_1^2 p_2^4 p_3^5 p_4^5 p_5^6+108 p_1^2 p_2^3 p_3^5 p_4^6 p_5^6) \textcolor{red}{{z}^{22}}+24 p_1^2 p_2^4 p_3^5 p_4^6 p_5^6 \textcolor{red}{{z}^{23}}+p_1^2 p_2^4 p_3^6 p_4^6 p_5^6 \textcolor{red}{{z}^{24}},
\)
\item[$H_4=$]\(
1+28 p_4 \textcolor{red}{z}+(168 p_3 p_4+210 p_4 p_5) \textcolor{red}{{z}^2}+ \dots
+(210 p_1^2 p_2^4 p_3^6 p_4^7 p_5^7+168 p_1^2 p_2^4 p_3^5 p_4^7 p_5^8) \textcolor{red}{{z}^{26}}+28 p_1^2 p_2^4 p_3^6 p_4^7 p_5^8 \textcolor{red}{{z}^{27}}+p_1^2 p_2^4 p_3^6 p_4^8 p_5^8 \textcolor{red}{{z}^{28}},
\)
\item[$H_5=$]\(
1+15 p_5 \textcolor{red}{z}+105 p_4 p_5 \textcolor{red}{{z}^2}+ \dots
+105 p_1 p_2^2 p_3^3 p_4^3 p_5^4 \textcolor{red}{{z}^{13}}+15 p_1 p_2^2 p_3^3 p_4^4 p_5^4 \textcolor{red}{{z}^{14}}+p_1 p_2^2 p_3^3 p_4^4 p_5^5 \textcolor{red}{{z}^{15}}.
\)
\end{itemize}

\bigskip 
\noindent { \bf $C_5$-case. }
 For the Lie algebra $C_5 $ the polynomials have the following structures 
\begin{itemize}
\item[$H_1=$]\(
1+9 p_1 \textcolor{red}{z}+36 p_1 p_2 \textcolor{red}{{z}^2}+ 
\dots +36 p_1 p_2 p_3^2 p_4^2 p_5 \textcolor{red}{{z}^7}+9 p_1 p_2^2 p_3^2 p_4^2 p_5 \textcolor{red}{{z}^8}+p_1^2 p_2^2 p_3^2 p_4^2 p_5 \textcolor{red}{{z}^9},
\)
\item[$H_2=$]\(
1+16 p_2 \textcolor{red}{z}+(36 p_1 p_2+84 p_2 p_3) \textcolor{red}{{z}^2}+ \dots
+(84 p_1^2 p_2^3 p_3^3 p_4^4 p_5^2+36 p_1 p_2^3 p_3^4 p_4^4 p_5^2) \textcolor{red}{{z}^{14}}+16 p_1^2 p_2^3 p_3^4 p_4^4 p_5^2 \textcolor{red}{{z}^{15}}+p_1^2 p_2^4 p_3^4 p_4^4 p_5^2 \textcolor{red}{{z}^{16}},
\)
\item[$H_3=$]\(
1+21 p_3 \textcolor{red}{z}+(84 p_2 p_3+126 p_3 p_4) \textcolor{red}{{z}^2}+ \dots +(126 p_1^2 p_2^4 p_3^5 p_4^5 p_5^3+84 p_1^2 p_2^3 p_3^5 p_4^6 p_5^3) \textcolor{red}{{z}^{19}}+21 p_1^2 p_2^4 p_3^5 p_4^6 p_5^3 \textcolor{red}{{z}^{20}}+p_1^2 p_2^4 p_3^6 p_4^6 p_5^3 \textcolor{red}{{z}^{21}},
\)
\item[$H_4=$]\(
1+24 p_4 \textcolor{red}{z}+(126 p_3 p_4+150 p_4 p_5) \textcolor{red}{{z}^2}+ \dots +(150 p_1^2 p_2^4 p_3^6 p_4^7 p_5^3+126 p_1^2 p_2^4 p_3^5 p_4^7 p_5^4) \textcolor{red}{{z}^{22}}+24 p_1^2 p_2^4 p_3^6 p_4^7 p_5^4 \textcolor{red}{{z}^{23}}+p_1^2 p_2^4 p_3^6 p_4^8 p_5^4 \textcolor{red}{{z}^{24}},
\)
\item[$H_5=$]\(
1+25 p_5 \textcolor{red}{z}+300 p_4 p_5 \textcolor{red}{{z}^2}+ \dots +300 p_1^2 p_2^4 p_3^6 p_4^7 p_5^4 \textcolor{red}{{z}^{23}}+25 p_1^2 p_2^4 p_3^6 p_4^8 p_5^4 \textcolor{red}{{z}^{24}}+p_1^2 p_2^4 p_3^6 p_4^8 p_5^5 \textcolor{red}{{z}^{25}}.
\)
\end{itemize}

\bigskip
\noindent { \bf $D_5$-case. }
 For the Lie algebra $D_5 $ we get the following structures of  polynomials 
\begin{itemize}
\item[$H_1=$]\(
1+8 p_1 \textcolor{red}{z}+28 p_1 p_2 \textcolor{red}{{z}^2}+ \dots +28 p_1 p_2 p_3^2 p_4 p_5 \textcolor{red}{{z}^6}+8 p_1 p_2^2 p_3^2 p_4 p_5 \textcolor{red}{{z}^7}+p_1^2 p_2^2 p_3^2 p_4 p_5 \textcolor{red}{{z}^8},
\)
\item[$H_2=$]\(
1+14 p_2 \textcolor{red}{z}+(28 p_1 p_2+63 p_2 p_3) \textcolor{red}{{z}^2}+ \dots +(63 p_1^2 p_2^3 p_3^3 p_4^2 p_5^2+28 p_1 p_2^3 p_3^4 p_4^2 p_5^2) \textcolor{red}{{z}^{12}}+14 p_1^2 p_2^3 p_3^4 p_4^2 p_5^2 \textcolor{red}{{z}^{13}}+p_1^2 p_2^4 p_3^4 p_4^2 p_5^2 \textcolor{red}{{z}^{14}},
\) %
\item[$H_3=$]\(
1+18 p_3 \textcolor{red}{z}+(63 p_2 p_3+45 p_3 p_4+45 p_3 p_5) \textcolor{red}{{z}^2}+ \dots +(45 p_1^2 p_2^4 p_3^5 p_4^3 p_5^2+45 p_1^2 p_2^4 p_3^5 p_4^2 p_5^3+63 p_1^2 p_2^3 p_3^5 p_4^3 p_5^3) \textcolor{red}{{z}^{16}}+18 p_1^2 p_2^4 p_3^5 p_4^3 p_5^3 \textcolor{red}{{z}^{17}}+p_1^2 p_2^4 p_3^6 p_4^3 p_5^3 \textcolor{red}{{z}^{18}},
\)
\item[$H_4=$]\(
1+10 p_4 \textcolor{red}{z}+45 p_3 p_4 \textcolor{red}{{z}^2}+ \dots +45 p_1 p_2^2 p_3^2 p_4^2 p_5 \textcolor{red}{{z}^8}+10 p_1 p_2^2 p_3^3 p_4^2 p_5 \textcolor{red}{{z}^9}+p_1 p_2^2 p_3^3 p_4^2 p_5^2 \textcolor{red}{{z}^{10}},
\)
\item[$H_5=$]\(
1+10 p_5 \textcolor{red}{z}+45 p_3 p_5 \textcolor{red}{{z}^2}+ \dots +45 p_1 p_2^2 p_3^2 p_4 p_5^2 \textcolor{red}{{z}^8}+10 p_1 p_2^2 p_3^3 p_4 p_5^2 \textcolor{red}{{z}^9}+p_1 p_2^2 p_3^3 p_4^2 p_5^2 \textcolor{red}{{z}^{10}}.
\)
\end{itemize}
     
Now we denote
      \begin{equation}
        \label{3e.5}
         H_s \equiv H_s(z) = H_s(z, (p_i) ), \quad (p_i) \equiv (p_1,p_2,p_3,p_4,p_5).
       \end{equation}
             
The polynomials have the following asymptotic behaviour

\begin{equation}
        \label{3e.6}
         H_s = H_s(z, (p_i) )  \sim \left( \prod_{l=1}^{5} (p_l)^{\nu^{sl}} \right) z^{n_s} \equiv 
         H_s^{as}(z, (p_i)), \quad \text{ as } z \to \infty,
       \end{equation}
     where  we denote by $\nu = (\nu^{sl})$ the integer valued matrix which  have the form 
\begin{align}
         \label{3e.7}
        \nu &=  
         \begin{pmatrix}
          1 & 1 & 1 & 1 & 1 \\
          1 & 2 & 2 & 2 & 1 \\
          1 & 2 & 3 & 2 & 1 \\
          1 & 2 & 2 & 2 & 1 \\
          1 & 1 & 1 & 1 & 1 \\
         \end{pmatrix},
         \quad  
         \begin{pmatrix}
           2 & 2 & 2 & 2 & 2 \\
           2 & 4 & 4 & 4 & 4 \\
           2 & 4 & 6 & 6 & 6 \\
           2 & 4 & 6 & 8 & 8 \\
           1 & 2 & 3 & 4 & 5 \\
         \end{pmatrix},
         \quad 
         \begin{pmatrix}
          2 & 2 & 2 & 2 & 1 \\
          2 & 4 & 4 & 4 & 2 \\
          2 & 4 & 6 & 6 & 3 \\
          2 & 4 & 6 & 8 & 4 \\
          2 & 4 & 6 & 8 & 5 \\
         \end{pmatrix},
         \quad \nonumber \\
        &\hspace{5em}
         \begin{pmatrix}
           2 & 2 & 2 & 1 & 1 \\
           2 & 4 & 4 & 2 & 2 \\
           2 & 4 & 6 & 3 & 3 \\
           1 & 2 & 3 & 2 & 2 \\
           1 & 2 & 3 & 2 & 2 \\
         \end{pmatrix}
\end{align}
      for Lie algebras $A_5,  B_5,  C_5, D_5$, respectively.
     It may be readily verified that the matrix $\nu = (\nu^{sl})$ obey 
     the following identity 
                   \begin{equation}
                     \label{3e.10a}
                      \sum_{l= 1}^5 \nu^{sl} = n_s, \quad s = 1, 2, 3, 4, 5.
                    \end{equation}
\\
It should be noted that for Lie algebras  $B_5$, $C_5$, 
 the $\nu$-matrix is coinciding with twice inverse Cartan matrix $A^{-1}$,
 i.e. 
      \begin{equation}
        \label{3e.8a}
      \nu({\cal G}) = 2 A^{-1}, \quad {\cal G}=B_5, C_5,
      \end{equation}
      while in the $A_5$ and $D_5$ cases   
      we have a more sophisticated relation
       \begin{equation}
        \label{3e.8b}
         \nu({\cal G})  = A^{-1} (I + P({\cal G})), \quad {\cal G}=A_5, D_5.
       \end{equation}
       Here we denote by  $I$  $5 \times 5$ identity matrix and by $P({\cal G})$ -
        a matrix corresponding to a certain permutation $\sigma \in S_5$ ($S_5$ is symmetric group) 
        by the relation: $P = (P^i_j) = (\delta^i_{\sigma(j)})$, 
       where $\sigma$ is the generator of the group 
       $G = \{ \sigma, {\rm id} \}$. $G$ is the group of symmetry 
       of the Dynkin diagram for $A_5$ and $D_5$  which act on the set 
       of corresponding five vertices via their permutations. In fact, the group $G$ is 
       isomorphic to the group $\mathbb{Z}_2$.      
Here we present explicit forms for the permutation matrix $P$ and the generator $\sigma$ for both Lie algebras 
$A_5, D_5$:
      \begin{equation}
         \label{3e.9a}
         P({A_5})  = 
         \begin{pmatrix}
           0 & 0 & 0 & 0 & 1 \\
           0 & 0 & 0 & 1 & 0 \\
           0 & 0 & 1 & 0 & 0 \\
           0 & 1 & 0 & 0 & 0 \\
           1 & 0 & 0 & 0 & 0 \\
         \end{pmatrix},\quad \sigma: (1,2,3,4,5) \mapsto (5,4,3,2,1);
         \end{equation}
      \begin{equation}
         \label{3e.9b}
         P({D_5})  = 
         \begin{pmatrix}
            1 & 0 & 0 & 0 & 0 \\
            0 & 1 & 0 & 0 & 0 \\
            0 & 0 & 1 & 0 & 0 \\
            0 & 0 & 0 & 0 & 1 \\
            0 & 0 & 0 & 1 & 0 \\
         \end{pmatrix},\quad \sigma: (1,2,3,4,5) \mapsto (1,2,3,5,4).
         \end{equation} \\ \indent       
    
    We note that the above symmetry groups control certain identity properties for polynomials $H_s(z)$.
  \\ \indent
         We denote $\hat{p}_i = p_{\sigma(i)} $ for the $A_5$ and $D_5$ cases, and
       $\hat{p}_i = p_{i}$ for $B_5$ and $C_5$  cases ($i= 1,2,3,4,5$).         
       The ordered set $(\hat{p}_i)$ is called  as \textit{dual} one to the ordered set $(p_i)$. 
       
      By using MATHEMATICA algorithms we verified the validity of the following identities.
   \\[1em] 
       {\bf Symmetry relations.}
   \\   
       {\bf Proposition 1.} {\em  
         The fluxbrane polynomials, corresponding to 
         Lie algebras for $A_5$ and $D_5$, obey for all $p_i$ and $z>0$
         the following identities:
        \bearr{3.11}
        H_{\sigma(s)}(z, (p_i) )\, = H_s(z, (\hat{p}_i)),
         \ear
         where $\sigma\in S_5$, $s= 1, \dots, 5$ is defined for each algebra by Eqs.
          \eqref{3e.9a}, \eqref{3e.9b}}.
         Relations (\ref{3.11}) may be called as symmetry ones.
   \\[1em]
       {\bf Duality relations.}
   \\  
       {\bf Proposition 2.} {\em  The fluxbrane polynomials which
        correspond to Lie algebras $A_5$,
        $B_5$, $C_5$, $D_5$, satisfy for all $p_i > 0$ and $z > 0$
        the following identities
        \begin{equation}
          \label{3.12}
           H_{s}(z, (p_i) ) = H_s^{as}(z, (p_i)) H_s(z^{-1}, (\hat{p}_i^{-1})),
         \end{equation}
         $s = 1, 2, 3, 4,5$. }                                                          
Relations (\ref{3.12}) may be called as duality ones.

   {\bf Fluxes.}
         Now we put our attention on oriented $2$-dimensional manifold 
        $M_{*} =(0, + \infty) \times S^1$. 
        We calculate the  flux integrals over this manifold:
         \beq{3.19}
          \Phi^s  = \int_{M_{*}} F^s =
           2 \pi \int_{0}^{ + \infty} d \rho \rho {\cal B}^s. 
         \eeq    
         Here 
          \begin{equation}
             \label{3e.16}
             {\cal B}^s =   q_s  \prod_{l = 1}^{5}  H_{l}^{- A_{s l}}.
          \end{equation} 
         \\
         Due to results of Ref. \cite{Ivas-flux-17} the  flux integrals $\Phi^s$ read
          \beq{3.25}
         \Phi^s =    4 \pi n_s q_s^{-1} h_s,  
        \eeq
     $s =1, 2,3,4,5$. 
     Here as in general case \cite{Ivas-flux-17} any flux $\Phi^s$ depends upon one integration constant  
     $q_s \neq 0$, while the integrand form $F^s$ depends upon all constants: $q_1, q_2, q_3, q_4,q_5$.

     We  note also that  by putting $q_1 = 0$ we get the Melvin-type solutions corresponding 
     to classical Lie algebras $A_4$,  $B_4$,  $C_4$,  $D_4$, respectively,  which were analyzed 
     in ref. \cite{Bol-Ivas-R4-18}. The case of rank 3 Lie algebras was considered in \cite{BolIvas-R3-18}.
     (For the case of the rank-2 Lie algebras see ref. \cite{BolIvas-R2-17}.)  
        
        {\bf Special solutions.} By putting         
                  $p_1 = p_2= p_3 = p_4 = p_5 = p > 0$ we get binomial relations
                  \beq{3.27}
                           H_{s}(z) =  H_{s}(z;(p,p,p,p,p)) = (1 + p z)^{n_s},        
                  \eeq
                 which obey  the master equations (\ref{1.1}) with boundary 
                 conditions (\ref{1.2}) imposed with parameters $q_s$ satisfying
                 the following relations 
                 \beq{3.27a}
                   \frac{1}{4} K_s q_s^2 / n_s =  p,        
                 \eeq
                 $s = 1,2,3,4,5$.
                 
       
         
         
         {\bf Asymptotic relations.}
           
           Now we present the asymptotic relations as $\rho \to + \infty $ for the solution under consideration :
                    \bear{3e.26}
                     g_{as} = \Bigl(\prod_{l = 1}^{5} p_l^{a_l} \Bigr)^{2/(D-2)} \rho^{2A}
                     \biggl\{ d\rho \otimes d \rho  \qquad \\ \nonumber
                      +
                     \Bigl(\prod_{l = 1}^{5} p_l^{a_l } \Bigr)^{-2} 
                     \rho^{2 - 2A (D-2)} d\phi \otimes d\phi +  g^2 \biggr\},
                     \\  \label{3e.27}
                     \varphi^a_{as}=  \sum_{s = 1}^{5} h_s \lambda_{s}^a 
                     (\sum_{l = 1}^{5} \nu^{sl} \ln p_l + 2 n_s \ln \rho ),
                     \\  \label{3e.28}
                     F^s_{as} = q_s  p_s^{-1} p_{\theta(s)}^{-1} \rho^{-3}  d\rho \wedge d \phi,
                     \ear       
                     $a, s =1,2,3,4,5$,
                     where 
                      \beq{3.25a}
                       a_l = \sum_{s =1}^5 h_s \nu^{sl}, \qquad   A =  2 (D-2)^{-1} \sum_{s = 1}^{5} n_{s} h_s, 
                      \eeq
                      and  in (\ref{3e.28}) we put $\theta = \sigma $ for ${\cal G} = A_5$, 
                  and $\theta = {\rm id} $ for ${\cal G} = B_5, C_5, D_5$.
                   
                    Now we  explain the appearance of these asymptotical relations.                              
                         Indeed, due to polynomial structure of moduli functions we have 
                        \beq{3e.20}
                         H_s \sim C_s \rho^{2n_s},   \qquad C_s = \prod_{l = 1}^{5} (p_l)^{\nu^{sl}},
                              \eeq 
                        as $\rho \to + \infty$. From (\ref{3e.16}), (\ref{3e.20}) and 
                        the equality $\sum_{1}^{n} A_{s l} n_l = 2$,
                        following from (\ref{1.4}), we get
                        \beq{3.21}
                        {\cal B}^s \sim  q_s C^s \rho^{-4},  \quad  
                         C^s = \prod_{l = 1}^{5}  p_{l}^{- (A \nu) _{s}^{ \ l}}.
                        \eeq 
                         $s =1,2,3,4,5$.   
                         
                          Using (\ref{3e.8b})  and (\ref{3.21}) we have for the $A_5$-case 
                            \beq{3e.23}
                                C^s =  \prod_{l = 1}^{5} p_l^{-(I+P)_{s}^{\ l}} = 
                               \prod_{l = 1}^{5} p_l^{- \delta^l_s - \delta^l_{\sigma(s)} } = 
                                     p_s^{-1} p_{\sigma(s)}^{-1}.               
                            \eeq 
                           Similarly, due to (\ref{3e.8a}) and  (\ref{3.21}) we get for Lie algebras $B_5$, $C_5$, $D_5$:            
                            \beq{3e.23cg}
                             C^s =  \prod_{l = 1}^{5} p_l^{- 2 \delta^l_s } =  p_s^{-2}. 
                            \eeq

           We note that   for ${\cal G} = B_5,  C_5, D_5$ 
          the asymptotic value of form
          $F^s_{as}$ depends upon $q_s$, $s = 1,2,3,4,5$. In the $A_5$-case  $F^s_{as}$ 
          depends: upon $q_1$ and $q_5$ for $s = 1,5$ and upon $q_2, q_4$
          for $s = 2,4$ and upon  $q_3$ for $s = 3$.

  \section{\bf Conclusions}
  
    In this paper, we have studied a family of generalized multidimensional Melvin-type solutions 
    which correspond to  simple Lie algebras of rank $5$:
    ${\cal G} = A_5,  B_5,  C_5, D_5$.  Any solution of this family is ruled by a set of $5$ 
    polynomials $H_s(z)$ of powers $n_s$, $s =1,2,3,4,5$. 
    The powers of these polynomials read: 
   $(n_1, n_2, n_3, n_4, n_5) = (5,8,9,8,5), (10,18,24,28,15), 
    (9,16,21,24,25), (8,14,18,10,10)$  
   for Lie algebras $A_5$, $B_5$, $C_5$, $D_5$ respectively.
    We have presented in Appendix all these 
    polynomials calculated by using a certain MATHEMATICA algorithm.  
    In fact, these (so-called fluxbrane) polynomials determine special 
    solutions to open Toda chain equations \cite{I-14} which correspond
    to the Lie algebras under consideration and may be used in various areas of 
    science.
         
    The moduli parameters $p_s$ of polynomials $H_s(z) = H_s(z,(p_s))$ 
    are related to  parameters $q_s$ by the relation $p_s = K_s q_s^2/(4 n_s)$, where
     $K_s$ depends upon the total dimension $D$ and dilatonic coupling vectors 
     $\vec{\lambda}_{s}$ by the relation (\ref{3.17}).
        For $D =4$ the  parameters $q_s$ determine (up to a sign $\pm$) 
     the values  of  colored magnetic fields on the axis of symmetry. 
     
    Here we have found the symmetry relations and the duality identities for our 
    rank-5 fluxbrane polynomials.  These identities may be also used 
   in deriving  $1/\rho$-expansion for solutions under consideration at large distances $\rho$, e.g. 
   for asymptotic relations (as  $\rho \to + \infty$) which are obtained in the paper.

   By using results of Ref. \cite{I-14} one can construct black hole solutions corresponding to 
   rank-5 Lie algebras for the model under consideration  along the lines of how it was done 
   in  \cite{Bol-Ivas-R4-18} for the rank-4 case.  In rank-5 case one will need a thorough analysis of horizons in black hole metrics governed by fluxbrane polynomials extended to negative values of variable $z$.  For  dyonic black hole solutions corresponding to rank-2 Lie algebras such analysis was started in fact in Refs. \cite{Dav,GalZad} (see also \cite{AIMT}). The proper analysis of black hole solutions corresponding to Lie algebras of ranks 3 and 4 is also desirable.    This may be a subject of our future papers.               
 
 \renewcommand{\theequation}{\Alph{section}.\arabic{equation}}
 \renewcommand{\thesection}{\Alph{section}}
 \setcounter{section}{0}
 
 \section{Appendix}
 
 \renewcommand{\thesubsection}{\Alph{section}}
 
 \subsection{The list of polynomials}
 
 \noindent{\bf $A_5$-case.} For the Lie algebra $A_5 \cong sl(6)$ the polynomials read
  \begin{itemize} 
 \item[$H_1=$]\(
 1+5 p_1 \textcolor{red}{z}+10 p_1 p_2 \textcolor{red}{{z}^2}+10 p_1 p_2 p_3 \textcolor{red}{{z}^3}+5 p_1 p_2 p_3 p_4 \textcolor{red}{{z}^4}+p_1 p_2 p_3 p_4 p_5 \textcolor{red}{{z}^5}
 \)
 \item[$H_2=$]\(
 1+8 p_2 \textcolor{red}{z}+(10 p_1 p_2+18 p_2 p_3) \textcolor{red}{{z}^2}+(40 p_1 p_2 p_3+16 p_2 p_3 p_4) \textcolor{red}{{z}^3}
 +(20 p_1 p_2^2 p_3+45 p_1 p_2 p_3 p_4+5 p_2 p_3 p_4 p_5) \textcolor{red}{{z}^4}+(40 p_1 p_2^2 p_3 p_4+16 p_1 p_2 p_3 p_4 p_5) \textcolor{red}{{z}^5}
 +(10 p_1 p_2^2 p_3^2 p_4+18 p_1 p_2^2 p_3 p_4 p_5) \textcolor{red}{{z}^6}+8 p_1 p_2^2 p_3^2 p_4 p_5 \textcolor{red}{{z}^7}+p_1 p_2^2 p_3^2 p_4^2 p_5 \textcolor{red}{{z}^8}\)
 \item[$H_3=$]\(
 1+9 p_3 \textcolor{red}{z}+(18 p_2 p_3+18 p_3 p_4) \textcolor{red}{{z}^2}+(10 p_1 p_2 p_3+64 p_2 p_3 p_4+10 p_3 p_4 p_5) \textcolor{red}{{z}^3}
 +(45 p_1 p_2 p_3 p_4+36 p_2 p_3^2 p_4+45 p_2 p_3 p_4 p_5) \textcolor{red}{{z}^4}+(45 p_1 p_2 p_3^2 p_4+36 p_1 p_2 p_3 p_4 p_5+45 p_2 p_3^2 p_4 p_5) \textcolor{red}{{z}^5}
 +(10 p_1 p_2^2 p_3^2 p_4+64 p_1 p_2 p_3^2 p_4 p_5+10 p_2 p_3^2 p_4^2 p_5) \textcolor{red}{{z}^6}+(18 p_1 p_2^2 p_3^2 p_4 p_5+18 p_1 p_2 p_3^2 p_4^2 p_5) \textcolor{red}{{z}^7}
 +9 p_1 p_2^2 p_3^2 p_4^2 p_5 \textcolor{red}{{z}^8}+p_1 p_2^2 p_3^3 p_4^2 p_5 \textcolor{red}{{z}^9}\)
 \item[$H_4=$]\(
 1+8 p_4 \textcolor{red}{z}+(18 p_3 p_4+10 p_4 p_5) \textcolor{red}{{z}^2}+(16 p_2 p_3 p_4+40 p_3 p_4 p_5) \textcolor{red}{{z}^3}
 +(5 p_1 p_2 p_3 p_4+45 p_2 p_3 p_4 p_5+20 p_3 p_4^2 p_5) \textcolor{red}{{z}^4}+(16 p_1 p_2 p_3 p_4 p_5+40 p_2 p_3 p_4^2 p_5) \textcolor{red}{{z}^5}
 +(18 p_1 p_2 p_3 p_4^2 p_5+10 p_2 p_3^2 p_4^2 p_5) \textcolor{red}{{z}^6}+8 p_1 p_2 p_3^2 p_4^2 p_5 \textcolor{red}{{z}^7}+p_1 p_2^2 p_3^2 p_4^2 p_5 \textcolor{red}{{z}^8}\)
 \item[$H_5=$]\(
 1+5 p_5 \textcolor{red}{z}+10 p_4 p_5 \textcolor{red}{{z}^2}+10 p_3 p_4 p_5 \textcolor{red}{{z}^3}+5 p_2 p_3 p_4 p_5 \textcolor{red}{{z}^4}+p_1 p_2 p_3 p_4 p_5 \textcolor{red}{{z}^5}
 \)
 \end{itemize}

 \bigskip
 \noindent {\bf $B_5$-case.}
 For the Lie algebra $B_5 \cong so(11)$ we obtain
 \begin{itemize}
 \item[$H_1=$]\(
 1+10 p_1 \textcolor{red}{z}+45 p_1 p_2 \textcolor{red}{{z}^2}+120 p_1 p_2 p_3 \textcolor{red}{{z}^3}+210 p_1 p_2 p_3 p_4 \textcolor{red}{{z}^4}+252 p_1 p_2 p_3 p_4 p_5 \textcolor{red}{{z}^5}+210 p_1 p_2 p_3 p_4 p_5^2 \textcolor{red}{{z}^6}
 +120 p_1 p_2 p_3 p_4^2 p_5^2 \textcolor{red}{{z}^7}+45 p_1 p_2 p_3^2 p_4^2 p_5^2 \textcolor{red}{{z}^8}+10 p_1 p_2^2 p_3^2 p_4^2 p_5^2 \textcolor{red}{{z}^9}+p_1^2 p_2^2 p_3^2 p_4^2 p_5^2 \textcolor{red}{{z}^{10}}
 \)
 \item[$H_2=$]\(
 1+18 p_2 \textcolor{red}{z}+(45 p_1 p_2+108 p_2 p_3) \textcolor{red}{{z}^2}+(480 p_1 p_2 p_3+336 p_2 p_3 p_4) \textcolor{red}{{z}^3}
 +(540 p_1 p_2^2 p_3+1890 p_1 p_2 p_3 p_4+630 p_2 p_3 p_4 p_5) \textcolor{red}{{z}^4}+(3780 p_1 p_2^2 p_3 p_4+4032 p_1 p_2 p_3 p_4 p_5 
 +756 p_2 p_3 p_4 p_5^2) \textcolor{red}{{z}^5}+(2520 p_1 p_2^2 p_3^2 p_4+10206 p_1 p_2^2 p_3 p_4 p_5+5250 p_1 p_2 p_3 p_4 p_5^2+588 p_2 p_3 p_4^2 p_5^2) \textcolor{red}{{z}^6}
 +(12096 p_1 p_2^2 p_3^2 p_4 p_5+15120 p_1 p_2^2 p_3 p_4 p_5^2+4320 p_1 p_2 p_3 p_4^2 p_5^2+288 p_2 p_3^2 p_4^2 p_5^2) \textcolor{red}{{z}^7}
 +(5292 p_1 p_2^2 p_3^2 p_4^2 p_5+22680 p_1 p_2^2 p_3^2 p_4 p_5^2+13500 p_1 p_2^2 p_3 p_4^2 p_5^2+2205 p_1 p_2 p_3^2 p_4^2 p_5^2+81 p_2^2 p_3^2 p_4^2 p_5^2) \textcolor{red}{{z}^8}
 +48620 p_1 p_2^2 p_3^2 p_4^2 p_5^2 \textcolor{red}{{z}^9}+(81 p_1^2 p_2^2 p_3^2 p_4^2 p_5^2+2205 p_1 p_2^3 p_3^2 p_4^2 p_5^2+13500 p_1 p_2^2 p_3^3 p_4^2 p_5^2
 +22680 p_1 p_2^2 p_3^2 p_4^3 p_5^2+5292 p_1 p_2^2 p_3^2 p_4^2 p_5^3) \textcolor{red}{{z}^{10}}+(288 p_1^2 p_2^3 p_3^2 p_4^2 p_5^2+4320 p_1 p_2^3 p_3^3 p_4^2 p_5^2
 +15120 p_1 p_2^2 p_3^3 p_4^3 p_5^2+12096 p_1 p_2^2 p_3^2 p_4^3 p_5^3) \textcolor{red}{{z}^{11}}+(588 p_1^2 p_2^3 p_3^3 p_4^2 p_5^2+5250 p_1 p_2^3 p_3^3 p_4^3 p_5^2
 +10206 p_1 p_2^2 p_3^3 p_4^3 p_5^3+2520 p_1 p_2^2 p_3^2 p_4^3 p_5^4) \textcolor{red}{{z}^{12}}+(756 p_1^2 p_2^3 p_3^3 p_4^3 p_5^2+4032 p_1 p_2^3 p_3^3 p_4^3 p_5^3
 +3780 p_1 p_2^2 p_3^3 p_4^3 p_5^4) \textcolor{red}{{z}^{13}}+(630 p_1^2 p_2^3 p_3^3 p_4^3 p_5^3+1890 p_1 p_2^3 p_3^3 p_4^3 p_5^4+540 p_1 p_2^2 p_3^3 p_4^4 p_5^4) \textcolor{red}{{z}^{14}}
 +(336 p_1^2 p_2^3 p_3^3 p_4^3 p_5^4+480 p_1 p_2^3 p_3^3 p_4^4 p_5^4) \textcolor{red}{{z}^{15}}+(108 p_1^2 p_2^3 p_3^3 p_4^4 p_5^4+45 p_1 p_2^3 p_3^4 p_4^4 p_5^4) \textcolor{red}{{z}^{16}}
 +18 p_1^2 p_2^3 p_3^4 p_4^4 p_5^4 \textcolor{red}{{z}^{17}}+p_1^2 p_2^4 p_3^4 p_4^4 p_5^4 \textcolor{red}{{z}^{18}}
 \)
 \item[$H_3=$]\(
 1+24 p_3 \textcolor{red}{z}+(108 p_2 p_3+168 p_3 p_4) \textcolor{red}{{z}^2}+(120 p_1 p_2 p_3+1344 p_2 p_3 p_4+560 p_3 p_4 p_5) \textcolor{red}{{z}^3}
 +(1890 p_1 p_2 p_3 p_4+2016 p_2 p_3^2 p_4+5670 p_2 p_3 p_4 p_5+1050 p_3 p_4 p_5^2) \textcolor{red}{{z}^4}
 +(5040 p_1 p_2 p_3^2 p_4+9072 p_1 p_2 p_3 p_4 p_5+15120 p_2 p_3^2 p_4 p_5+12096 p_2 p_3 p_4 p_5^2+1176 p_3 p_4^2 p_5^2) \textcolor{red}{{z}^5}
 +(2520 p_1 p_2^2 p_3^2 p_4+43008 p_1 p_2 p_3^2 p_4 p_5+11760 p_2 p_3^2 p_4^2 p_5+21000 p_1 p_2 p_3 p_4 p_5^2
 +40824 p_2 p_3^2 p_4 p_5^2+14700 p_2 p_3 p_4^2 p_5^2+784 p_3^2 p_4^2 p_5^2) \textcolor{red}{{z}^6}+(27216 p_1 p_2^2 p_3^2 p_4 p_5
 +42336 p_1 p_2 p_3^2 p_4^2 p_5+126000 p_1 p_2 p_3^2 p_4 p_5^2+27000 p_1 p_2 p_3 p_4^2 p_5^2+123552 p_2 p_3^2 p_4^2 p_5^2) \textcolor{red}{{z}^7}
 +(47628 p_1 p_2^2 p_3^2 p_4^2 p_5+90720 p_1 p_2^2 p_3^2 p_4 p_5^2+424710 p_1 p_2 p_3^2 p_4^2 p_5^2
 +3969 p_2^2 p_3^2 p_4^2 p_5^2+43200 p_2 p_3^3 p_4^2 p_5^2+98784 p_2 p_3^2 p_4^3 p_5^2+26460 p_2 p_3^2 p_4^2 p_5^3) \textcolor{red}{{z}^8}
 +(14112 p_1 p_2^2 p_3^3 p_4^2 p_5+434720 p_1 p_2^2 p_3^2 p_4^2 p_5^2+147000 p_1 p_2 p_3^3 p_4^2 p_5^2+17496 p_2^2 p_3^3 p_4^2 p_5^2
 +408240 p_1 p_2 p_3^2 p_4^3 p_5^2+86016 p_2 p_3^3 p_4^3 p_5^2+117600 p_1 p_2 p_3^2 p_4^2 p_5^3+82320 p_2 p_3^2 p_4^3 p_5^3) \textcolor{red}{{z}^9}
 +(1296 p_1^2 p_2^2 p_3^2 p_4^2 p_5^2+291720 p_1 p_2^2 p_3^3 p_4^2 p_5^2+567000 p_1 p_2^2 p_3^2 p_4^3 p_5^2
 +370440 p_1 p_2 p_3^3 p_4^3 p_5^2+37800 p_2^2 p_3^3 p_4^3 p_5^2+190512 p_1 p_2^2 p_3^2 p_4^2 p_5^3+387072 p_1 p_2 p_3^2 p_4^3 p_5^3
 +90720 p_2 p_3^3 p_4^3 p_5^3+24696 p_2 p_3^2 p_4^3 p_5^4) \textcolor{red}{{z}^{10}}+(10584 p_1^2 p_2^2 p_3^3 p_4^2 p_5^2+52920 p_1 p_2^3 p_3^3 p_4^2 p_5^2
 +960960 p_1 p_2^2 p_3^3 p_4^3 p_5^2+127008 p_1 p_2^2 p_3^3 p_4^2 p_5^3+680400 p_1 p_2^2 p_3^2 p_4^3 p_5^3
 +444528 p_1 p_2 p_3^3 p_4^3 p_5^3+45360 p_2^2 p_3^3 p_4^3 p_5^3+126000 p_1 p_2 p_3^2 p_4^3 p_5^4+48384 p_2 p_3^3 p_4^3 p_5^4) \textcolor{red}{{z}^{11}}
 +(9408 p_1^2 p_2^3 p_3^3 p_4^2 p_5^2+30618 p_1^2 p_2^2 p_3^3 p_4^3 p_5^2+257250 p_1 p_2^3 p_3^3 p_4^3 p_5^2+252000 p_1 p_2^2 p_3^4 p_4^3 p_5^2
 +1605604 p_1 p_2^2 p_3^3 p_4^3 p_5^3+252000 p_1 p_2^2 p_3^2 p_4^3 p_5^4+257250 p_1 p_2 p_3^3 p_4^3 p_5^4
 +30618 p_2^2 p_3^3 p_4^3 p_5^4+9408 p_2 p_3^3 p_4^4 p_5^4) \textcolor{red}{{z}^{12}}+(48384 p_1^2 p_2^3 p_3^3 p_4^3 p_5^2
 +126000 p_1 p_2^3 p_3^4 p_4^3 p_5^2+45360 p_1^2 p_2^2 p_3^3 p_4^3 p_5^3+444528 p_1 p_2^3 p_3^3 p_4^3 p_5^3
 +680400 p_1 p_2^2 p_3^4 p_4^3 p_5^3+127008 p_1 p_2^2 p_3^3 p_4^4 p_5^3+960960 p_1 p_2^2 p_3^3 p_4^3 p_5^4
 +52920 p_1 p_2 p_3^3 p_4^4 p_5^4+10584 p_2^2 p_3^3 p_4^4 p_5^4) \textcolor{red}{{z}^{13}}+(24696 p_1^2 p_2^3 p_3^4 p_4^3 p_5^2+90720 p_1^2 p_2^3 p_3^3 p_4^3 p_5^3+387072 p_1 p_2^3 p_3^4 p_4^3 p_5^3+190512 p_1 p_2^2 p_3^4 p_4^4 p_5^3+37800 p_1^2 p_2^2 p_3^3 p_4^3 p_5^4+370440 p_1 p_2^3 p_3^3 p_4^3 p_5^4+567000 p_1 p_2^2 p_3^4 p_4^3 p_5^4+291720 p_1 p_2^2 p_3^3 p_4^4 p_5^4+1296 p_2^2 p_3^4 p_4^4 p_5^4) \textcolor{red}{{z}^{14}}+(82320 p_1^2 p_2^3 p_3^4 p_4^3 p_5^3+117600 p_1 p_2^3 p_3^4 p_4^4 p_5^3+86016 p_1^2 p_2^3 p_3^3 p_4^3 p_5^4+408240 p_1 p_2^3 p_3^4 p_4^3 p_5^4+17496 p_1^2 p_2^2 p_3^3 p_4^4 p_5^4+147000 p_1 p_2^3 p_3^3 p_4^4 p_5^4+434720 p_1 p_2^2 p_3^4 p_4^4 p_5^4+14112 p_1 p_2^2 p_3^3 p_4^4 p_5^5) \textcolor{red}{{z}^{15}}+(26460 p_1^2 p_2^3 p_3^4 p_4^4 p_5^3+98784 p_1^2 p_2^3 p_3^4 p_4^3 p_5^4+43200 p_1^2 p_2^3 p_3^3 p_4^4 p_5^4+3969 p_1^2 p_2^2 p_3^4 p_4^4 p_5^4+424710 p_1 p_2^3 p_3^4 p_4^4 p_5^4+90720 p_1 p_2^2 p_3^4 p_4^5 p_5^4+47628 p_1 p_2^2 p_3^4 p_4^4 p_5^5) \textcolor{red}{{z}^{16}}+(123552 p_1^2 p_2^3 p_3^4 p_4^4 p_5^4+27000 p_1 p_2^3 p_3^5 p_4^4 p_5^4+126000 p_1 p_2^3 p_3^4 p_4^5 p_5^4+42336 p_1 p_2^3 p_3^4 p_4^4 p_5^5+27216 p_1 p_2^2 p_3^4 p_4^5 p_5^5) \textcolor{red}{{z}^{17}}+(784 p_1^2 p_2^4 p_3^4 p_4^4 p_5^4+14700 p_1^2 p_2^3 p_3^5 p_4^4 p_5^4+40824 p_1^2 p_2^3 p_3^4 p_4^5 p_5^4+21000 p_1 p_2^3 p_3^5 p_4^5 p_5^4+11760 p_1^2 p_2^3 p_3^4 p_4^4 p_5^5+43008 p_1 p_2^3 p_3^4 p_4^5 p_5^5+2520 p_1 p_2^2 p_3^4 p_4^5 p_5^6) \textcolor{red}{{z}^{18}}+(1176 p_1^2 p_2^4 p_3^5 p_4^4 p_5^4+12096 p_1^2 p_2^3 p_3^5 p_4^5 p_5^4+15120 p_1^2 p_2^3 p_3^4 p_4^5 p_5^5+9072 p_1 p_2^3 p_3^5 p_4^5 p_5^5+5040 p_1 p_2^3 p_3^4 p_4^5 p_5^6) \textcolor{red}{{z}^{19}}+(1050 p_1^2 p_2^4 p_3^5 p_4^5 p_5^4+5670 p_1^2 p_2^3 p_3^5 p_4^5 p_5^5+2016 p_1^2 p_2^3 p_3^4 p_4^5 p_5^6+1890 p_1 p_2^3 p_3^5 p_4^5 p_5^6) \textcolor{red}{{z}^{20}}+(560 p_1^2 p_2^4 p_3^5 p_4^5 p_5^5+1344 p_1^2 p_2^3 p_3^5 p_4^5 p_5^6+120 p_1 p_2^3 p_3^5 p_4^6 p_5^6) \textcolor{red}{{z}^{21}}+(168 p_1^2 p_2^4 p_3^5 p_4^5 p_5^6+108 p_1^2 p_2^3 p_3^5 p_4^6 p_5^6) \textcolor{red}{{z}^{22}}+24 p_1^2 p_2^4 p_3^5 p_4^6 p_5^6 \textcolor{red}{{z}^{23}}+p_1^2 p_2^4 p_3^6 p_4^6 p_5^6 \textcolor{red}{{z}^{24}}
 \)
 \item[$H_4=$]\(
 1+28 p_4 \textcolor{red}{z}+(168 p_3 p_4+210 p_4 p_5) \textcolor{red}{{z}^2}+(336 p_2 p_3 p_4+2240 p_3 p_4 p_5+700 p_4 p_5^2) \textcolor{red}{{z}^3}
 +(210 p_1 p_2 p_3 p_4+5670 p_2 p_3 p_4 p_5+3920 p_3 p_4^2 p_5+9450 p_3 p_4 p_5^2+1225 p_4^2 p_5^2) \textcolor{red}{{z}^4}+(4032 p_1 p_2 p_3 p_4 p_5+17640 p_2 p_3 p_4^2 p_5+27216 p_2 p_3 p_4 p_5^2+49392 p_3 p_4^2 p_5^2) \textcolor{red}{{z}^5}+(15876 p_1 p_2 p_3 p_4^2 p_5+11760 p_2 p_3^2 p_4^2 p_5+21000 p_1 p_2 p_3 p_4 p_5^2+209916 p_2 p_3 p_4^2 p_5^2+19600 p_3^2 p_4^2 p_5^2+74088 p_3 p_4^3 p_5^2+24500 p_3 p_4^2 p_5^3) \textcolor{red}{{z}^6}+(18816 p_1 p_2 p_3^2 p_4^2 p_5+195120 p_1 p_2 p_3 p_4^2 p_5^2+202176 p_2 p_3^2 p_4^2 p_5^2+411600 p_2 p_3 p_4^3 p_5^2+87808 p_3^2 p_4^3 p_5^2+158760 p_2 p_3 p_4^2 p_5^3+109760 p_3 p_4^3 p_5^3) \textcolor{red}{{z}^7}+(5292 p_1 p_2^2 p_3^2 p_4^2 p_5+277830 p_1 p_2 p_3^2 p_4^2 p_5^2+35721 p_2^2 p_3^2 p_4^2 p_5^2+425250 p_1 p_2 p_3 p_4^3 p_5^2+961632 p_2 p_3^2 p_4^3 p_5^2+176400 p_1 p_2 p_3 p_4^2 p_5^3+238140 p_2 p_3^2 p_4^2 p_5^3+771750 p_2 p_3 p_4^3 p_5^3+164640 p_3^2 p_4^3 p_5^3+51450 p_3 p_4^3 p_5^4) \textcolor{red}{{z}^8}+(109760 p_1 p_2^2 p_3^2 p_4^2 p_5^2+1292760 p_1 p_2 p_3^2 p_4^3 p_5^2+308700 p_2^2 p_3^2 p_4^3 p_5^2+537600 p_2 p_3^3 p_4^3 p_5^2+470400 p_1 p_2 p_3^2 p_4^2 p_5^3+907200 p_1 p_2 p_3 p_4^3 p_5^3+2731680 p_2 p_3^2 p_4^3 p_5^3+411600 p_2 p_3 p_4^3 p_5^4+137200 p_3^2 p_4^3 p_5^4) \textcolor{red}{{z}^9}+(7056 p_1^2 p_2^2 p_3^2 p_4^2 p_5^2+666680 p_1 p_2^2 p_3^2 p_4^3 p_5^2+1029000 p_1 p_2 p_3^3 p_4^3 p_5^2+340200 p_2^2 p_3^3 p_4^3 p_5^2+190512 p_1 p_2^2 p_3^2 p_4^2 p_5^3+4484844 p_1 p_2 p_3^2 p_4^3 p_5^3+833490 p_2^2 p_3^2 p_4^3 p_5^3+2268000 p_2 p_3^3 p_4^3 p_5^3+576240 p_2 p_3^2 p_4^4 p_5^3+525000 p_1 p_2 p_3 p_4^3 p_5^4+2163672 p_2 p_3^2 p_4^3 p_5^4+38416 p_3^2 p_4^4 p_5^4) \textcolor{red}{{z}^{10}}+(81648 p_1^2 p_2^2 p_3^2 p_4^3 p_5^2+1132320 p_1 p_2^2 p_3^3 p_4^3 p_5^2+2621472 p_1 p_2^2 p_3^2 p_4^3 p_5^3+4939200 p_1 p_2 p_3^3 p_4^3 p_5^3+1632960 p_2^2 p_3^3 p_4^3 p_5^3+1524096 p_1 p_2 p_3^2 p_4^4 p_5^3+1128960 p_2 p_3^3 p_4^4 p_5^3+3591000 p_1 p_2 p_3^2 p_4^3 p_5^4+1000188 p_2^2 p_3^2 p_4^3 p_5^4+2721600 p_2 p_3^3 p_4^3 p_5^4+1100736 p_2 p_3^2 p_4^4 p_5^4) \textcolor{red}{{z}^{11}}+(166698 p_1^2 p_2^2 p_3^3 p_4^3 p_5^2+257250 p_1 p_2^3 p_3^3 p_4^3 p_5^2+272160 p_1^2 p_2^2 p_3^2 p_4^3 p_5^3+6419812 p_1 p_2^2 p_3^3 p_4^3 p_5^3+1190700 p_1 p_2^2 p_3^2 p_4^4 p_5^3+3111696 p_1 p_2 p_3^3 p_4^4 p_5^3+882000 p_2^2 p_3^3 p_4^4 p_5^3+2666720 p_1 p_2^2 p_3^2 p_4^3 p_5^4+6431250 p_1 p_2 p_3^3 p_4^3 p_5^4+2480058 p_2^2 p_3^3 p_4^3 p_5^4+2500470 p_1 p_2 p_3^2 p_4^4 p_5^4+540225 p_2^2 p_3^2 p_4^4 p_5^4+3358656 p_2 p_3^3 p_4^4 p_5^4+144060 p_2 p_3^2 p_4^4 p_5^5) \textcolor{red}{{z}^{12}}+(65856 p_1^2 p_2^3 p_3^3 p_4^3 p_5^2+987840 p_1^2 p_2^2 p_3^3 p_4^3 p_5^3+1778112 p_1 p_2^3 p_3^3 p_4^3 p_5^3+5551504 p_1 p_2^2 p_3^3 p_4^4 p_5^3+403200 p_1^2 p_2^2 p_3^2 p_4^3 p_5^4+10190880 p_1 p_2^2 p_3^3 p_4^3 p_5^4+2744000 p_1 p_2^2 p_3^2 p_4^4 p_5^4+9560880 p_1 p_2 p_3^3 p_4^4 p_5^4+4000752 p_2^2 p_3^3 p_4^4 p_5^4+1053696 p_2 p_3^3 p_4^5 p_5^4+470400 p_1 p_2 p_3^2 p_4^4 p_5^5+635040 p_2 p_3^3 p_4^4 p_5^5) \textcolor{red}{{z}^{13}}+(493920 p_1^2 p_2^3 p_3^3 p_4^3 p_5^3+714420 p_1^2 p_2^2 p_3^3 p_4^4 p_5^3+2160900 p_1 p_2^3 p_3^3 p_4^4 p_5^3+529200 p_1 p_2^2 p_3^4 p_4^4 p_5^3+1852200 p_1^2 p_2^2 p_3^3 p_4^3 p_5^4+3333960 p_1 p_2^3 p_3^3 p_4^3 p_5^4+291600 p_1^2 p_2^2 p_3^2 p_4^4 p_5^4+21364200 p_1 p_2^2 p_3^3 p_4^4 p_5^4+291600 p_2^2 p_3^4 p_4^4 p_5^4+3333960 p_1 p_2 p_3^3 p_4^5 p_5^4+1852200 p_2^2 p_3^3 p_4^5 p_5^4+529200 p_1 p_2^2 p_3^2 p_4^4 p_5^5+2160900 p_1 p_2 p_3^3 p_4^4 p_5^5+714420 p_2^2 p_3^3 p_4^4 p_5^5+493920 p_2 p_3^3 p_4^5 p_5^5) \textcolor{red}{{z}^{14}}+(635040 p_1^2 p_2^3 p_3^3 p_4^4 p_5^3+470400 p_1 p_2^3 p_3^4 p_4^4 p_5^3+1053696 p_1^2 p_2^3 p_3^3 p_4^3 p_5^4+4000752 p_1^2 p_2^2 p_3^3 p_4^4 p_5^4+9560880 p_1 p_2^3 p_3^3 p_4^4 p_5^4+2744000 p_1 p_2^2 p_3^4 p_4^4 p_5^4+10190880 p_1 p_2^2 p_3^3 p_4^5 p_5^4+403200 p_2^2 p_3^4 p_4^5 p_5^4+5551504 p_1 p_2^2 p_3^3 p_4^4 p_5^5+1778112 p_1 p_2 p_3^3 p_4^5 p_5^5+987840 p_2^2 p_3^3 p_4^5 p_5^5+65856 p_2 p_3^3 p_4^5 p_5^6) \textcolor{red}{{z}^{15}}+(144060 p_1^2 p_2^3 p_3^4 p_4^4 p_5^3+3358656 p_1^2 p_2^3 p_3^3 p_4^4 p_5^4+540225 p_1^2 p_2^2 p_3^4 p_4^4 p_5^4+2500470 p_1 p_2^3 p_3^4 p_4^4 p_5^4+2480058 p_1^2 p_2^2 p_3^3 p_4^5 p_5^4+6431250 p_1 p_2^3 p_3^3 p_4^5 p_5^4+2666720 p_1 p_2^2 p_3^4 p_4^5 p_5^4+882000 p_1^2 p_2^2 p_3^3 p_4^4 p_5^5+3111696 p_1 p_2^3 p_3^3 p_4^4 p_5^5+1190700 p_1 p_2^2 p_3^4 p_4^4 p_5^5+6419812 p_1 p_2^2 p_3^3 p_4^5 p_5^5+272160 p_2^2 p_3^4 p_4^5 p_5^5+257250 p_1 p_2 p_3^3 p_4^5 p_5^6+166698 p_2^2 p_3^3 p_4^5 p_5^6) \textcolor{red}{{z}^{16}}+(1100736 p_1^2 p_2^3 p_3^4 p_4^4 p_5^4+2721600 p_1^2 p_2^3 p_3^3 p_4^5 p_5^4+1000188 p_1^2 p_2^2 p_3^4 p_4^5 p_5^4+3591000 p_1 p_2^3 p_3^4 p_4^5 p_5^4+1128960 p_1^2 p_2^3 p_3^3 p_4^4 p_5^5+1524096 p_1 p_2^3 p_3^4 p_4^4 p_5^5+1632960 p_1^2 p_2^2 p_3^3 p_4^5 p_5^5+4939200 p_1 p_2^3 p_3^3 p_4^5 p_5^5+2621472 p_1 p_2^2 p_3^4 p_4^5 p_5^5+1132320 p_1 p_2^2 p_3^3 p_4^5 p_5^6+81648 p_2^2 p_3^4 p_4^5 p_5^6) \textcolor{red}{{z}^{17}}+(38416 p_1^2 p_2^4 p_3^4 p_4^4 p_5^4+2163672 p_1^2 p_2^3 p_3^4 p_4^5 p_5^4+525000 p_1 p_2^3 p_3^5 p_4^5 p_5^4+576240 p_1^2 p_2^3 p_3^4 p_4^4 p_5^5+2268000 p_1^2 p_2^3 p_3^3 p_4^5 p_5^5+833490 p_1^2 p_2^2 p_3^4 p_4^5 p_5^5+4484844 p_1 p_2^3 p_3^4 p_4^5 p_5^5+190512 p_1 p_2^2 p_3^4 p_4^6 p_5^5+340200 p_1^2 p_2^2 p_3^3 p_4^5 p_5^6+1029000 p_1 p_2^3 p_3^3 p_4^5 p_5^6+666680 p_1 p_2^2 p_3^4 p_4^5 p_5^6+7056 p_2^2 p_3^4 p_4^6 p_5^6) \textcolor{red}{{z}^{18}}+(137200 p_1^2 p_2^4 p_3^4 p_4^5 p_5^4+411600 p_1^2 p_2^3 p_3^5 p_4^5 p_5^4+2731680 p_1^2 p_2^3 p_3^4 p_4^5 p_5^5+907200 p_1 p_2^3 p_3^5 p_4^5 p_5^5+470400 p_1 p_2^3 p_3^4 p_4^6 p_5^5+537600 p_1^2 p_2^3 p_3^3 p_4^5 p_5^6+308700 p_1^2 p_2^2 p_3^4 p_4^5 p_5^6+1292760 p_1 p_2^3 p_3^4 p_4^5 p_5^6+109760 p_1 p_2^2 p_3^4 p_4^6 p_5^6) \textcolor{red}{{z}^{19}}+(51450 p_1^2 p_2^4 p_3^5 p_4^5 p_5^4+164640 p_1^2 p_2^4 p_3^4 p_4^5 p_5^5+771750 p_1^2 p_2^3 p_3^5 p_4^5 p_5^5+238140 p_1^2 p_2^3 p_3^4 p_4^6 p_5^5+176400 p_1 p_2^3 p_3^5 p_4^6 p_5^5+961632 p_1^2 p_2^3 p_3^4 p_4^5 p_5^6+425250 p_1 p_2^3 p_3^5 p_4^5 p_5^6+35721 p_1^2 p_2^2 p_3^4 p_4^6 p_5^6+277830 p_1 p_2^3 p_3^4 p_4^6 p_5^6+5292 p_1 p_2^2 p_3^4 p_4^6 p_5^7) \textcolor{red}{{z}^{20}}+(109760 p_1^2 p_2^4 p_3^5 p_4^5 p_5^5+158760 p_1^2 p_2^3 p_3^5 p_4^6 p_5^5+87808 p_1^2 p_2^4 p_3^4 p_4^5 p_5^6+411600 p_1^2 p_2^3 p_3^5 p_4^5 p_5^6+202176 p_1^2 p_2^3 p_3^4 p_4^6 p_5^6+195120 p_1 p_2^3 p_3^5 p_4^6 p_5^6+18816 p_1 p_2^3 p_3^4 p_4^6 p_5^7) \textcolor{red}{{z}^{21}}+(24500 p_1^2 p_2^4 p_3^5 p_4^6 p_5^5+74088 p_1^2 p_2^4 p_3^5 p_4^5 p_5^6+19600 p_1^2 p_2^4 p_3^4 p_4^6 p_5^6+209916 p_1^2 p_2^3 p_3^5 p_4^6 p_5^6+21000 p_1 p_2^3 p_3^5 p_4^7 p_5^6+11760 p_1^2 p_2^3 p_3^4 p_4^6 p_5^7+15876 p_1 p_2^3 p_3^5 p_4^6 p_5^7) \textcolor{red}{{z}^{22}}+(49392 p_1^2 p_2^4 p_3^5 p_4^6 p_5^6+27216 p_1^2 p_2^3 p_3^5 p_4^7 p_5^6+17640 p_1^2 p_2^3 p_3^5 p_4^6 p_5^7+4032 p_1 p_2^3 p_3^5 p_4^7 p_5^7) \textcolor{red}{{z}^{23}}+(1225 p_1^2 p_2^4 p_3^6 p_4^6 p_5^6+9450 p_1^2 p_2^4 p_3^5 p_4^7 p_5^6+3920 p_1^2 p_2^4 p_3^5 p_4^6 p_5^7+5670 p_1^2 p_2^3 p_3^5 p_4^7 p_5^7+210 p_1 p_2^3 p_3^5 p_4^7 p_5^8) \textcolor{red}{{z}^{24}}+(700 p_1^2 p_2^4 p_3^6 p_4^7 p_5^6+2240 p_1^2 p_2^4 p_3^5 p_4^7 p_5^7+336 p_1^2 p_2^3 p_3^5 p_4^7 p_5^8) \textcolor{red}{{z}^{25}}+(210 p_1^2 p_2^4 p_3^6 p_4^7 p_5^7+168 p_1^2 p_2^4 p_3^5 p_4^7 p_5^8) \textcolor{red}{{z}^{26}}+28 p_1^2 p_2^4 p_3^6 p_4^7 p_5^8 \textcolor{red}{{z}^{27}}+p_1^2 p_2^4 p_3^6 p_4^8 p_5^8 \textcolor{red}{{z}^{28}}
 \)
 \item[$H_5=$]\(
 1+15 p_5 \textcolor{red}{z}+105 p_4 p_5 \textcolor{red}{{z}^2}+(280 p_3 p_4 p_5+175 p_4 p_5^2) \textcolor{red}{{z}^3}+(315 p_2 p_3 p_4 p_5+1050 p_3 p_4 p_5^2) \textcolor{red}{{z}^4}+(126 p_1 p_2 p_3 p_4 p_5+1701 p_2 p_3 p_4 p_5^2+1176 p_3 p_4^2 p_5^2) \textcolor{red}{{z}^5}+(840 p_1 p_2 p_3 p_4 p_5^2+3675 p_2 p_3 p_4^2 p_5^2+490 p_3 p_4^2 p_5^3) \textcolor{red}{{z}^6}+(2430 p_1 p_2 p_3 p_4^2 p_5^2+1800 p_2 p_3^2 p_4^2 p_5^2+2205 p_2 p_3 p_4^2 p_5^3) \textcolor{red}{{z}^7}+(2205 p_1 p_2 p_3^2 p_4^2 p_5^2+1800 p_1 p_2 p_3 p_4^2 p_5^3+2430 p_2 p_3^2 p_4^2 p_5^3) \textcolor{red}{{z}^8}+(490 p_1 p_2^2 p_3^2 p_4^2 p_5^2+3675 p_1 p_2 p_3^2 p_4^2 p_5^3+840 p_2 p_3^2 p_4^3 p_5^3) \textcolor{red}{{z}^9}+(1176 p_1 p_2^2 p_3^2 p_4^2 p_5^3+1701 p_1 p_2 p_3^2 p_4^3 p_5^3+126 p_2 p_3^2 p_4^3 p_5^4) \textcolor{red}{{z}^{10}}+(1050 p_1 p_2^2 p_3^2 p_4^3 p_5^3+315 p_1 p_2 p_3^2 p_4^3 p_5^4) \textcolor{red}{{z}^{11}}+(175 p_1 p_2^2 p_3^3 p_4^3 p_5^3+280 p_1 p_2^2 p_3^2 p_4^3 p_5^4) \textcolor{red}{{z}^{12}}+105 p_1 p_2^2 p_3^3 p_4^3 p_5^4 \textcolor{red}{{z}^{13}}+15 p_1 p_2^2 p_3^3 p_4^4 p_5^4 \textcolor{red}{{z}^{14}}+p_1 p_2^2 p_3^3 p_4^4 p_5^5 \textcolor{red}{{z}^{15}}
 \)
 \end{itemize}

 \bigskip 
 \noindent { \bf $C_5$-case. }
 For the Lie algebra $C_5 \cong sp(5)$ we find 
 \begin{itemize}
 \item[$H_1=$]\(
 1+9 p_1 \textcolor{red}{z}+36 p_1 p_2 \textcolor{red}{{z}^2}+84 p_1 p_2 p_3 \textcolor{red}{{z}^3}+126 p_1 p_2 p_3 p_4 \textcolor{red}{{z}^4}+126 p_1 p_2 p_3 p_4 p_5 \textcolor{red}{{z}^5}+84 p_1 p_2 p_3 p_4^2 p_5 \textcolor{red}{{z}^6}+36 p_1 p_2 p_3^2 p_4^2 p_5 \textcolor{red}{{z}^7}+9 p_1 p_2^2 p_3^2 p_4^2 p_5 \textcolor{red}{{z}^8}+p_1^2 p_2^2 p_3^2 p_4^2 p_5 \textcolor{red}{{z}^9}
 \)
 \item[$H_2=$]\(
 1+16 p_2 \textcolor{red}{z}+(36 p_1 p_2+84 p_2 p_3) \textcolor{red}{{z}^2}+(336 p_1 p_2 p_3+224 p_2 p_3 p_4) \textcolor{red}{{z}^3}+(336 p_1 p_2^2 p_3+1134 p_1 p_2 p_3 p_4+350 p_2 p_3 p_4 p_5) \textcolor{red}{{z}^4}+(2016 p_1 p_2^2 p_3 p_4+2016 p_1 p_2 p_3 p_4 p_5+336 p_2 p_3 p_4^2 p_5) \textcolor{red}{{z}^5}+(1176 p_1 p_2^2 p_3^2 p_4+4536 p_1 p_2^2 p_3 p_4 p_5+2100 p_1 p_2 p_3 p_4^2 p_5+196 p_2 p_3^2 p_4^2 p_5) \textcolor{red}{{z}^6}+(4704 p_1 p_2^2 p_3^2 p_4 p_5+5376 p_1 p_2^2 p_3 p_4^2 p_5+1296 p_1 p_2 p_3^2 p_4^2 p_5+64 p_2^2 p_3^2 p_4^2 p_5) \textcolor{red}{{z}^7}+12870 p_1 p_2^2 p_3^2 p_4^2 p_5 \textcolor{red}{{z}^8}+(64 p_1^2 p_2^2 p_3^2 p_4^2 p_5+1296 p_1 p_2^3 p_3^2 p_4^2 p_5+5376 p_1 p_2^2 p_3^3 p_4^2 p_5+4704 p_1 p_2^2 p_3^2 p_4^3 p_5) \textcolor{red}{{z}^9}+(196 p_1^2 p_2^3 p_3^2 p_4^2 p_5+2100 p_1 p_2^3 p_3^3 p_4^2 p_5+4536 p_1 p_2^2 p_3^3 p_4^3 p_5+1176 p_1 p_2^2 p_3^2 p_4^3 p_5^2) \textcolor{red}{{z}^{10}}+(336 p_1^2 p_2^3 p_3^3 p_4^2 p_5+2016 p_1 p_2^3 p_3^3 p_4^3 p_5+2016 p_1 p_2^2 p_3^3 p_4^3 p_5^2) \textcolor{red}{{z}^{11}}+(350 p_1^2 p_2^3 p_3^3 p_4^3 p_5+1134 p_1 p_2^3 p_3^3 p_4^3 p_5^2+336 p_1 p_2^2 p_3^3 p_4^4 p_5^2) \textcolor{red}{{z}^{12}}+(224 p_1^2 p_2^3 p_3^3 p_4^3 p_5^2+336 p_1 p_2^3 p_3^3 p_4^4 p_5^2) \textcolor{red}{{z}^{13}}+(84 p_1^2 p_2^3 p_3^3 p_4^4 p_5^2+36 p_1 p_2^3 p_3^4 p_4^4 p_5^2) \textcolor{red}{{z}^{14}}+16 p_1^2 p_2^3 p_3^4 p_4^4 p_5^2 \textcolor{red}{{z}^{15}}+p_1^2 p_2^4 p_3^4 p_4^4 p_5^2 \textcolor{red}{{z}^{16}}
 \)
 \item[$H_3=$]\(
 1+21 p_3 \textcolor{red}{z}+(84 p_2 p_3+126 p_3 p_4) \textcolor{red}{{z}^2}+(84 p_1 p_2 p_3+896 p_2 p_3 p_4+350 p_3 p_4 p_5) \textcolor{red}{{z}^3}+(1134 p_1 p_2 p_3 p_4+1176 p_2 p_3^2 p_4+3150 p_2 p_3 p_4 p_5+525 p_3 p_4^2 p_5) \textcolor{red}{{z}^4}+(2646 p_1 p_2 p_3^2 p_4+4536 p_1 p_2 p_3 p_4 p_5+7350 p_2 p_3^2 p_4 p_5+5376 p_2 p_3 p_4^2 p_5+441 p_3^2 p_4^2 p_5) \textcolor{red}{{z}^5}+(1176 p_1 p_2^2 p_3^2 p_4+18816 p_1 p_2 p_3^2 p_4 p_5+8400 p_1 p_2 p_3 p_4^2 p_5+25872 p_2 p_3^2 p_4^2 p_5) \textcolor{red}{{z}^6}+(10584 p_1 p_2^2 p_3^2 p_4 p_5+68112 p_1 p_2 p_3^2 p_4^2 p_5+2304 p_2^2 p_3^2 p_4^2 p_5+16464 p_2 p_3^3 p_4^2 p_5+18816 p_2 p_3^2 p_4^3 p_5) \textcolor{red}{{z}^7}+(48510 p_1 p_2^2 p_3^2 p_4^2 p_5+48384 p_1 p_2 p_3^3 p_4^2 p_5+8400 p_2^2 p_3^3 p_4^2 p_5+66150 p_1 p_2 p_3^2 p_4^3 p_5+24696 p_2 p_3^3 p_4^3 p_5+7350 p_2 p_3^2 p_4^3 p_5^2) \textcolor{red}{{z}^8}+(784 p_1^2 p_2^2 p_3^2 p_4^2 p_5+65142 p_1 p_2^2 p_3^3 p_4^2 p_5+75264 p_1 p_2^2 p_3^2 p_4^3 p_5+91854 p_1 p_2 p_3^3 p_4^3 p_5+14336 p_2^2 p_3^3 p_4^3 p_5+29400 p_1 p_2 p_3^2 p_4^3 p_5^2+17150 p_2 p_3^3 p_4^3 p_5^2) \textcolor{red}{{z}^9}+(5376 p_1^2 p_2^2 p_3^3 p_4^2 p_5+18900 p_1 p_2^3 p_3^3 p_4^2 p_5+196812 p_1 p_2^2 p_3^3 p_4^3 p_5+42336 p_1 p_2^2 p_3^2 p_4^3 p_5^2+72576 p_1 p_2 p_3^3 p_4^3 p_5^2+12600 p_2^2 p_3^3 p_4^3 p_5^2+4116 p_2 p_3^3 p_4^4 p_5^2) \textcolor{red}{{z}^{10}}+(4116 p_1^2 p_2^3 p_3^3 p_4^2 p_5+12600 p_1^2 p_2^2 p_3^3 p_4^3 p_5+72576 p_1 p_2^3 p_3^3 p_4^3 p_5+42336 p_1 p_2^2 p_3^4 p_4^3 p_5+196812 p_1 p_2^2 p_3^3 p_4^3 p_5^2+18900 p_1 p_2 p_3^3 p_4^4 p_5^2+5376 p_2^2 p_3^3 p_4^4 p_5^2) \textcolor{red}{{z}^{11}}+(17150 p_1^2 p_2^3 p_3^3 p_4^3 p_5+29400 p_1 p_2^3 p_3^4 p_4^3 p_5+14336 p_1^2 p_2^2 p_3^3 p_4^3 p_5^2+91854 p_1 p_2^3 p_3^3 p_4^3 p_5^2+75264 p_1 p_2^2 p_3^4 p_4^3 p_5^2+65142 p_1 p_2^2 p_3^3 p_4^4 p_5^2+784 p_2^2 p_3^4 p_4^4 p_5^2) \textcolor{red}{{z}^{12}}+(7350 p_1^2 p_2^3 p_3^4 p_4^3 p_5+24696 p_1^2 p_2^3 p_3^3 p_4^3 p_5^2+66150 p_1 p_2^3 p_3^4 p_4^3 p_5^2+8400 p_1^2 p_2^2 p_3^3 p_4^4 p_5^2+48384 p_1 p_2^3 p_3^3 p_4^4 p_5^2+48510 p_1 p_2^2 p_3^4 p_4^4 p_5^2) \textcolor{red}{{z}^{13}}+(18816 p_1^2 p_2^3 p_3^4 p_4^3 p_5^2+16464 p_1^2 p_2^3 p_3^3 p_4^4 p_5^2+2304 p_1^2 p_2^2 p_3^4 p_4^4 p_5^2+68112 p_1 p_2^3 p_3^4 p_4^4 p_5^2+10584 p_1 p_2^2 p_3^4 p_4^5 p_5^2) \textcolor{red}{{z}^{14}}+(25872 p_1^2 p_2^3 p_3^4 p_4^4 p_5^2+8400 p_1 p_2^3 p_3^5 p_4^4 p_5^2+18816 p_1 p_2^3 p_3^4 p_4^5 p_5^2+1176 p_1 p_2^2 p_3^4 p_4^5 p_5^3) \textcolor{red}{{z}^{15}}+(441 p_1^2 p_2^4 p_3^4 p_4^4 p_5^2+5376 p_1^2 p_2^3 p_3^5 p_4^4 p_5^2+7350 p_1^2 p_2^3 p_3^4 p_4^5 p_5^2+4536 p_1 p_2^3 p_3^5 p_4^5 p_5^2+2646 p_1 p_2^3 p_3^4 p_4^5 p_5^3) \textcolor{red}{{z}^{16}}+(525 p_1^2 p_2^4 p_3^5 p_4^4 p_5^2+3150 p_1^2 p_2^3 p_3^5 p_4^5 p_5^2+1176 p_1^2 p_2^3 p_3^4 p_4^5 p_5^3+1134 p_1 p_2^3 p_3^5 p_4^5 p_5^3) \textcolor{red}{{z}^{17}}+(350 p_1^2 p_2^4 p_3^5 p_4^5 p_5^2+896 p_1^2 p_2^3 p_3^5 p_4^5 p_5^3+84 p_1 p_2^3 p_3^5 p_4^6 p_5^3) \textcolor{red}{{z}^{18}}+(126 p_1^2 p_2^4 p_3^5 p_4^5 p_5^3+84 p_1^2 p_2^3 p_3^5 p_4^6 p_5^3) \textcolor{red}{{z}^{19}}+21 p_1^2 p_2^4 p_3^5 p_4^6 p_5^3 \textcolor{red}{{z}^{20}}+p_1^2 p_2^4 p_3^6 p_4^6 p_5^3 \textcolor{red}{{z}^{21}}
 \)
 \item[$H_4=$]\(
 1+24 p_4 \textcolor{red}{z}+(126 p_3 p_4+150 p_4 p_5) \textcolor{red}{{z}^2}+(224 p_2 p_3 p_4+1400 p_3 p_4 p_5+400 p_4^2 p_5) \textcolor{red}{{z}^3}+(126 p_1 p_2 p_3 p_4+3150 p_2 p_3 p_4 p_5+7350 p_3 p_4^2 p_5) \textcolor{red}{{z}^4}+(2016 p_1 p_2 p_3 p_4 p_5+20832 p_2 p_3 p_4^2 p_5+7056 p_3^2 p_4^2 p_5+12600 p_3 p_4^3 p_5) \textcolor{red}{{z}^5}+(15288 p_1 p_2 p_3 p_4^2 p_5+29400 p_2 p_3^2 p_4^2 p_5+57344 p_2 p_3 p_4^3 p_5+23814 p_3^2 p_4^3 p_5+8750 p_3 p_4^3 p_5^2) \textcolor{red}{{z}^6}+(22752 p_1 p_2 p_3^2 p_4^2 p_5+14400 p_2^2 p_3^2 p_4^2 p_5+50400 p_1 p_2 p_3 p_4^3 p_5+178752 p_2 p_3^2 p_4^3 p_5+50400 p_2 p_3 p_4^3 p_5^2+29400 p_3^2 p_4^3 p_5^2) \textcolor{red}{{z}^7}+(16758 p_1 p_2^2 p_3^2 p_4^2 p_5+180900 p_1 p_2 p_3^2 p_4^3 p_5+98304 p_2^2 p_3^2 p_4^3 p_5+98784 p_2 p_3^3 p_4^3 p_5+50400 p_1 p_2 p_3 p_4^3 p_5^2+279300 p_2 p_3^2 p_4^3 p_5^2+11025 p_3^2 p_4^4 p_5^2) \textcolor{red}{{z}^8}+(3136 p_1^2 p_2^2 p_3^2 p_4^2 p_5+143472 p_1 p_2^2 p_3^2 p_4^3 p_5+163296 p_1 p_2 p_3^3 p_4^3 p_5+89600 p_2^2 p_3^3 p_4^3 p_5+321600 p_1 p_2 p_3^2 p_4^3 p_5^2+194400 p_2^2 p_3^2 p_4^3 p_5^2+274400 p_2 p_3^3 p_4^3 p_5^2+117600 p_2 p_3^2 p_4^4 p_5^2) \textcolor{red}{{z}^9}+(29400 p_1^2 p_2^2 p_3^2 p_4^3 p_5+233100 p_1 p_2^2 p_3^3 p_4^3 p_5+322812 p_1 p_2^2 p_3^2 p_4^3 p_5^2+516096 p_1 p_2 p_3^3 p_4^3 p_5^2+315000 p_2^2 p_3^3 p_4^3 p_5^2+142200 p_1 p_2 p_3^2 p_4^4 p_5^2+147456 p_2^2 p_3^2 p_4^4 p_5^2+255192 p_2 p_3^3 p_4^4 p_5^2) \textcolor{red}{{z}^{10}}+(50400 p_1^2 p_2^2 p_3^3 p_4^3 p_5+50400 p_1 p_2^3 p_3^3 p_4^3 p_5+75264 p_1^2 p_2^2 p_3^2 p_4^3 p_5^2+932400 p_1 p_2^2 p_3^3 p_4^3 p_5^2+268128 p_1 p_2^2 p_3^2 p_4^4 p_5^2+550368 p_1 p_2 p_3^3 p_4^4 p_5^2+470400 p_2^2 p_3^3 p_4^4 p_5^2+98784 p_2 p_3^3 p_4^5 p_5^2) \textcolor{red}{{z}^{11}}+(17150 p_1^2 p_2^3 p_3^3 p_4^3 p_5+229376 p_1^2 p_2^2 p_3^3 p_4^3 p_5^2+255150 p_1 p_2^3 p_3^3 p_4^3 p_5^2+78400 p_1^2 p_2^2 p_3^2 p_4^4 p_5^2+1544004 p_1 p_2^2 p_3^3 p_4^4 p_5^2+78400 p_2^2 p_3^4 p_4^4 p_5^2+255150 p_1 p_2 p_3^3 p_4^5 p_5^2+229376 p_2^2 p_3^3 p_4^5 p_5^2+17150 p_2 p_3^3 p_4^5 p_5^3) \textcolor{red}{{z}^{12}}+(98784 p_1^2 p_2^3 p_3^3 p_4^3 p_5^2+470400 p_1^2 p_2^2 p_3^3 p_4^4 p_5^2+550368 p_1 p_2^3 p_3^3 p_4^4 p_5^2+268128 p_1 p_2^2 p_3^4 p_4^4 p_5^2+932400 p_1 p_2^2 p_3^3 p_4^5 p_5^2+75264 p_2^2 p_3^4 p_4^5 p_5^2+50400 p_1 p_2 p_3^3 p_4^5 p_5^3+50400 p_2^2 p_3^3 p_4^5 p_5^3) \textcolor{red}{{z}^{13}}+(255192 p_1^2 p_2^3 p_3^3 p_4^4 p_5^2+147456 p_1^2 p_2^2 p_3^4 p_4^4 p_5^2+142200 p_1 p_2^3 p_3^4 p_4^4 p_5^2+315000 p_1^2 p_2^2 p_3^3 p_4^5 p_5^2+516096 p_1 p_2^3 p_3^3 p_4^5 p_5^2+322812 p_1 p_2^2 p_3^4 p_4^5 p_5^2+233100 p_1 p_2^2 p_3^3 p_4^5 p_5^3+29400 p_2^2 p_3^4 p_4^5 p_5^3) \textcolor{red}{{z}^{14}}+(117600 p_1^2 p_2^3 p_3^4 p_4^4 p_5^2+274400 p_1^2 p_2^3 p_3^3 p_4^5 p_5^2+194400 p_1^2 p_2^2 p_3^4 p_4^5 p_5^2+321600 p_1 p_2^3 p_3^4 p_4^5 p_5^2+89600 p_1^2 p_2^2 p_3^3 p_4^5 p_5^3+163296 p_1 p_2^3 p_3^3 p_4^5 p_5^3+143472 p_1 p_2^2 p_3^4 p_4^5 p_5^3+3136 p_2^2 p_3^4 p_4^6 p_5^3) \textcolor{red}{{z}^{15}}+(11025 p_1^2 p_2^4 p_3^4 p_4^4 p_5^2+279300 p_1^2 p_2^3 p_3^4 p_4^5 p_5^2+50400 p_1 p_2^3 p_3^5 p_4^5 p_5^2+98784 p_1^2 p_2^3 p_3^3 p_4^5 p_5^3+98304 p_1^2 p_2^2 p_3^4 p_4^5 p_5^3+180900 p_1 p_2^3 p_3^4 p_4^5 p_5^3+16758 p_1 p_2^2 p_3^4 p_4^6 p_5^3) \textcolor{red}{{z}^{16}}+(29400 p_1^2 p_2^4 p_3^4 p_4^5 p_5^2+50400 p_1^2 p_2^3 p_3^5 p_4^5 p_5^2+178752 p_1^2 p_2^3 p_3^4 p_4^5 p_5^3+50400 p_1 p_2^3 p_3^5 p_4^5 p_5^3+14400 p_1^2 p_2^2 p_3^4 p_4^6 p_5^3+22752 p_1 p_2^3 p_3^4 p_4^6 p_5^3) \textcolor{red}{{z}^{17}}+(8750 p_1^2 p_2^4 p_3^5 p_4^5 p_5^2+23814 p_1^2 p_2^4 p_3^4 p_4^5 p_5^3+57344 p_1^2 p_2^3 p_3^5 p_4^5 p_5^3+29400 p_1^2 p_2^3 p_3^4 p_4^6 p_5^3+15288 p_1 p_2^3 p_3^5 p_4^6 p_5^3) \textcolor{red}{{z}^{18}}+(12600 p_1^2 p_2^4 p_3^5 p_4^5 p_5^3+7056 p_1^2 p_2^4 p_3^4 p_4^6 p_5^3+20832 p_1^2 p_2^3 p_3^5 p_4^6 p_5^3+2016 p_1 p_2^3 p_3^5 p_4^7 p_5^3) \textcolor{red}{{z}^{19}}+(7350 p_1^2 p_2^4 p_3^5 p_4^6 p_5^3+3150 p_1^2 p_2^3 p_3^5 p_4^7 p_5^3+126 p_1 p_2^3 p_3^5 p_4^7 p_5^4) \textcolor{red}{{z}^{20}}+(400 p_1^2 p_2^4 p_3^6 p_4^6 p_5^3+1400 p_1^2 p_2^4 p_3^5 p_4^7 p_5^3+224 p_1^2 p_2^3 p_3^5 p_4^7 p_5^4) \textcolor{red}{{z}^{21}}+(150 p_1^2 p_2^4 p_3^6 p_4^7 p_5^3+126 p_1^2 p_2^4 p_3^5 p_4^7 p_5^4) \textcolor{red}{{z}^{22}}+24 p_1^2 p_2^4 p_3^6 p_4^7 p_5^4 \textcolor{red}{{z}^{23}}+p_1^2 p_2^4 p_3^6 p_4^8 p_5^4 \textcolor{red}{{z}^{24}}
 \)
 \item[$H_5=$]\(
 1+25 p_5 \textcolor{red}{z}+300 p_4 p_5 \textcolor{red}{{z}^2}+(700 p_3 p_4 p_5+1600 p_4^2 p_5) \textcolor{red}{{z}^3}+(700 p_2 p_3 p_4 p_5+9450 p_3 p_4^2 p_5+2500 p_4^2 p_5^2) \textcolor{red}{{z}^4}+(252 p_1 p_2 p_3 p_4 p_5+10752 p_2 p_3 p_4^2 p_5+15876 p_3^2 p_4^2 p_5+26250 p_3 p_4^2 p_5^2) \textcolor{red}{{z}^5}+(4200 p_1 p_2 p_3 p_4^2 p_5+39200 p_2 p_3^2 p_4^2 p_5+37800 p_2 p_3 p_4^2 p_5^2+78400 p_3^2 p_4^2 p_5^2+17500 p_3 p_4^3 p_5^2) \textcolor{red}{{z}^6}+(16200 p_1 p_2 p_3^2 p_4^2 p_5+25600 p_2^2 p_3^2 p_4^2 p_5+16800 p_1 p_2 p_3 p_4^2 p_5^2+245000 p_2 p_3^2 p_4^2 p_5^2+44800 p_2 p_3 p_4^3 p_5^2+132300 p_3^2 p_4^3 p_5^2) \textcolor{red}{{z}^7}+(22050 p_1 p_2^2 p_3^2 p_4^2 p_5+115200 p_1 p_2 p_3^2 p_4^2 p_5^2+202500 p_2^2 p_3^2 p_4^2 p_5^2+25200 p_1 p_2 p_3 p_4^3 p_5^2+617400 p_2 p_3^2 p_4^3 p_5^2+99225 p_3^2 p_4^4 p_5^2) \textcolor{red}{{z}^8}+(4900 p_1^2 p_2^2 p_3^2 p_4^2 p_5+198450 p_1 p_2^2 p_3^2 p_4^2 p_5^2+353400 p_1 p_2 p_3^2 p_4^3 p_5^2+691200 p_2^2 p_3^2 p_4^3 p_5^2+137200 p_2 p_3^3 p_4^3 p_5^2+627200 p_2 p_3^2 p_4^4 p_5^2+30625 p_3^2 p_4^4 p_5^3) \textcolor{red}{{z}^9}+(50176 p_1^2 p_2^2 p_3^2 p_4^2 p_5^2+798504 p_1 p_2^2 p_3^2 p_4^3 p_5^2+145152 p_1 p_2 p_3^3 p_4^3 p_5^2+280000 p_2^2 p_3^3 p_4^3 p_5^2+405000 p_1 p_2 p_3^2 p_4^4 p_5^2+1048576 p_2^2 p_3^2 p_4^4 p_5^2+296352 p_2 p_3^3 p_4^4 p_5^2+245000 p_2 p_3^2 p_4^4 p_5^3) \textcolor{red}{{z}^{10}}+(235200 p_1^2 p_2^2 p_3^2 p_4^3 p_5^2+491400 p_1 p_2^2 p_3^3 p_4^3 p_5^2+1411200 p_1 p_2^2 p_3^2 p_4^4 p_5^2+340200 p_1 p_2 p_3^3 p_4^4 p_5^2+1075200 p_2^2 p_3^3 p_4^4 p_5^2+180000 p_1 p_2 p_3^2 p_4^4 p_5^3+518400 p_2^2 p_3^2 p_4^4 p_5^3+205800 p_2 p_3^3 p_4^4 p_5^3) \textcolor{red}{{z}^{11}}+(179200 p_1^2 p_2^2 p_3^3 p_4^3 p_5^2+56700 p_1 p_2^3 p_3^3 p_4^3 p_5^2+490000 p_1^2 p_2^2 p_3^2 p_4^4 p_5^2+2118900 p_1 p_2^2 p_3^3 p_4^4 p_5^2+313600 p_2^2 p_3^4 p_4^4 p_5^2+793800 p_1 p_2^2 p_3^2 p_4^4 p_5^3+268800 p_1 p_2 p_3^3 p_4^4 p_5^3+945000 p_2^2 p_3^3 p_4^4 p_5^3+34300 p_2 p_3^3 p_4^5 p_5^3) \textcolor{red}{{z}^{12}}+(34300 p_1^2 p_2^3 p_3^3 p_4^3 p_5^2+945000 p_1^2 p_2^2 p_3^3 p_4^4 p_5^2+268800 p_1 p_2^3 p_3^3 p_4^4 p_5^2+793800 p_1 p_2^2 p_3^4 p_4^4 p_5^2+313600 p_1^2 p_2^2 p_3^2 p_4^4 p_5^3+2118900 p_1 p_2^2 p_3^3 p_4^4 p_5^3+490000 p_2^2 p_3^4 p_4^4 p_5^3+56700 p_1 p_2 p_3^3 p_4^5 p_5^3+179200 p_2^2 p_3^3 p_4^5 p_5^3) \textcolor{red}{{z}^{13}}+(205800 p_1^2 p_2^3 p_3^3 p_4^4 p_5^2+518400 p_1^2 p_2^2 p_3^4 p_4^4 p_5^2+180000 p_1 p_2^3 p_3^4 p_4^4 p_5^2+1075200 p_1^2 p_2^2 p_3^3 p_4^4 p_5^3+340200 p_1 p_2^3 p_3^3 p_4^4 p_5^3+1411200 p_1 p_2^2 p_3^4 p_4^4 p_5^3+491400 p_1 p_2^2 p_3^3 p_4^5 p_5^3+235200 p_2^2 p_3^4 p_4^5 p_5^3) \textcolor{red}{{z}^{14}}+(245000 p_1^2 p_2^3 p_3^4 p_4^4 p_5^2+296352 p_1^2 p_2^3 p_3^3 p_4^4 p_5^3+1048576 p_1^2 p_2^2 p_3^4 p_4^4 p_5^3+405000 p_1 p_2^3 p_3^4 p_4^4 p_5^3+280000 p_1^2 p_2^2 p_3^3 p_4^5 p_5^3+145152 p_1 p_2^3 p_3^3 p_4^5 p_5^3+798504 p_1 p_2^2 p_3^4 p_4^5 p_5^3+50176 p_2^2 p_3^4 p_4^6 p_5^3) \textcolor{red}{{z}^{15}}+(30625 p_1^2 p_2^4 p_3^4 p_4^4 p_5^2+627200 p_1^2 p_2^3 p_3^4 p_4^4 p_5^3+137200 p_1^2 p_2^3 p_3^3 p_4^5 p_5^3+691200 p_1^2 p_2^2 p_3^4 p_4^5 p_5^3+353400 p_1 p_2^3 p_3^4 p_4^5 p_5^3+198450 p_1 p_2^2 p_3^4 p_4^6 p_5^3+4900 p_2^2 p_3^4 p_4^6 p_5^4) \textcolor{red}{{z}^{16}}+(99225 p_1^2 p_2^4 p_3^4 p_4^4 p_5^3+617400 p_1^2 p_2^3 p_3^4 p_4^5 p_5^3+25200 p_1 p_2^3 p_3^5 p_4^5 p_5^3+202500 p_1^2 p_2^2 p_3^4 p_4^6 p_5^3+115200 p_1 p_2^3 p_3^4 p_4^6 p_5^3+22050 p_1 p_2^2 p_3^4 p_4^6 p_5^4) \textcolor{red}{{z}^{17}}+(132300 p_1^2 p_2^4 p_3^4 p_4^5 p_5^3+44800 p_1^2 p_2^3 p_3^5 p_4^5 p_5^3+245000 p_1^2 p_2^3 p_3^4 p_4^6 p_5^3+16800 p_1 p_2^3 p_3^5 p_4^6 p_5^3+25600 p_1^2 p_2^2 p_3^4 p_4^6 p_5^4+16200 p_1 p_2^3 p_3^4 p_4^6 p_5^4) \textcolor{red}{{z}^{18}}+(17500 p_1^2 p_2^4 p_3^5 p_4^5 p_5^3+78400 p_1^2 p_2^4 p_3^4 p_4^6 p_5^3+37800 p_1^2 p_2^3 p_3^5 p_4^6 p_5^3+39200 p_1^2 p_2^3 p_3^4 p_4^6 p_5^4+4200 p_1 p_2^3 p_3^5 p_4^6 p_5^4) \textcolor{red}{{z}^{19}}+(26250 p_1^2 p_2^4 p_3^5 p_4^6 p_5^3+15876 p_1^2 p_2^4 p_3^4 p_4^6 p_5^4+10752 p_1^2 p_2^3 p_3^5 p_4^6 p_5^4+252 p_1 p_2^3 p_3^5 p_4^7 p_5^4) \textcolor{red}{{z}^{20}}+(2500 p_1^2 p_2^4 p_3^6 p_4^6 p_5^3+9450 p_1^2 p_2^4 p_3^5 p_4^6 p_5^4+700 p_1^2 p_2^3 p_3^5 p_4^7 p_5^4) \textcolor{red}{{z}^{21}}+(1600 p_1^2 p_2^4 p_3^6 p_4^6 p_5^4+700 p_1^2 p_2^4 p_3^5 p_4^7 p_5^4) \textcolor{red}{{z}^{22}}+300 p_1^2 p_2^4 p_3^6 p_4^7 p_5^4 \textcolor{red}{{z}^{23}}+25 p_1^2 p_2^4 p_3^6 p_4^8 p_5^4 \textcolor{red}{{z}^{24}}+p_1^2 p_2^4 p_3^6 p_4^8 p_5^5 \textcolor{red}{{z}^{25}}
 \)
 \end{itemize}

 \bigskip
 \noindent { \bf $D_5$-case. }
  For the Lie algebra $D_5 \cong so(10)$ we find the following  polynomials
 \begin{itemize}
 \item[$H_1=$]\(
 1+8 p_1 \textcolor{red}{z}+28 p_1 p_2 \textcolor{red}{{z}^2}+56 p_1 p_2 p_3 \textcolor{red}{{z}^3}+(35 p_1 p_2 p_3 p_4+35 p_1 p_2 p_3 p_5) \textcolor{red}{{z}^4}+56 p_1 p_2 p_3 p_4 p_5 \textcolor{red}{{z}^5}+28 p_1 p_2 p_3^2 p_4 p_5 \textcolor{red}{{z}^6}+8 p_1 p_2^2 p_3^2 p_4 p_5 \textcolor{red}{{z}^7}+p_1^2 p_2^2 p_3^2 p_4 p_5 \textcolor{red}{{z}^8}
 \)
 \item[$H_2=$]\(
 1+14 p_2 \textcolor{red}{z}+(28 p_1 p_2+63 p_2 p_3) \textcolor{red}{{z}^2}+(224 p_1 p_2 p_3+70 p_2 p_3 p_4+70 p_2 p_3 p_5) \textcolor{red}{{z}^3}+(196 p_1 p_2^2 p_3+315 p_1 p_2 p_3 p_4+315 p_1 p_2 p_3 p_5+175 p_2 p_3 p_4 p_5) \textcolor{red}{{z}^4}+(490 p_1 p_2^2 p_3 p_4+490 p_1 p_2^2 p_3 p_5+896 p_1 p_2 p_3 p_4 p_5+126 p_2 p_3^2 p_4 p_5) \textcolor{red}{{z}^5}+(245 p_1 p_2^2 p_3^2 p_4+245 p_1 p_2^2 p_3^2 p_5+1764 p_1 p_2^2 p_3 p_4 p_5+700 p_1 p_2 p_3^2 p_4 p_5+49 p_2^2 p_3^2 p_4 p_5) \textcolor{red}{{z}^6}+3432 p_1 p_2^2 p_3^2 p_4 p_5 \textcolor{red}{{z}^7}+(49 p_1^2 p_2^2 p_3^2 p_4 p_5+700 p_1 p_2^3 p_3^2 p_4 p_5+1764 p_1 p_2^2 p_3^3 p_4 p_5+245 p_1 p_2^2 p_3^2 p_4^2 p_5+245 p_1 p_2^2 p_3^2 p_4 p_5^2) \textcolor{red}{{z}^8}+(126 p_1^2 p_2^3 p_3^2 p_4 p_5+896 p_1 p_2^3 p_3^3 p_4 p_5+490 p_1 p_2^2 p_3^3 p_4^2 p_5+490 p_1 p_2^2 p_3^3 p_4 p_5^2) \textcolor{red}{{z}^9}+(175 p_1^2 p_2^3 p_3^3 p_4 p_5+315 p_1 p_2^3 p_3^3 p_4^2 p_5+315 p_1 p_2^3 p_3^3 p_4 p_5^2+196 p_1 p_2^2 p_3^3 p_4^2 p_5^2) \textcolor{red}{{z}^{10}}+(70 p_1^2 p_2^3 p_3^3 p_4^2 p_5+70 p_1^2 p_2^3 p_3^3 p_4 p_5^2+224 p_1 p_2^3 p_3^3 p_4^2 p_5^2) \textcolor{red}{{z}^{11}}+(63 p_1^2 p_2^3 p_3^3 p_4^2 p_5^2+28 p_1 p_2^3 p_3^4 p_4^2 p_5^2) \textcolor{red}{{z}^{12}}+14 p_1^2 p_2^3 p_3^4 p_4^2 p_5^2 \textcolor{red}{{z}^{13}}+p_1^2 p_2^4 p_3^4 p_4^2 p_5^2 \textcolor{red}{{z}^{14}}
 \) %
 \item[$H_3=$]\(
 1+18 p_3 \textcolor{red}{z}+(63 p_2 p_3+45 p_3 p_4+45 p_3 p_5) \textcolor{red}{{z}^2}+(56 p_1 p_2 p_3+280 p_2 p_3 p_4+280 p_2 p_3 p_5+200 p_3 p_4 p_5) \textcolor{red}{{z}^3}+(315 p_1 p_2 p_3 p_4+315 p_2 p_3^2 p_4+315 p_1 p_2 p_3 p_5+315 p_2 p_3^2 p_5+1575 p_2 p_3 p_4 p_5+225 p_3^2 p_4 p_5) \textcolor{red}{{z}^4}+(630 p_1 p_2 p_3^2 p_4+630 p_1 p_2 p_3^2 p_5+2016 p_1 p_2 p_3 p_4 p_5+5292 p_2 p_3^2 p_4 p_5) \textcolor{red}{{z}^5}+(245 p_1 p_2^2 p_3^2 p_4+245 p_1 p_2^2 p_3^2 p_5+9996 p_1 p_2 p_3^2 p_4 p_5+1225 p_2^2 p_3^2 p_4 p_5+5103 p_2 p_3^3 p_4 p_5+875 p_2 p_3^2 p_4^2 p_5+875 p_2 p_3^2 p_4 p_5^2) \textcolor{red}{{z}^6}+(5616 p_1 p_2^2 p_3^2 p_4 p_5+12600 p_1 p_2 p_3^3 p_4 p_5+3528 p_2^2 p_3^3 p_4 p_5+2520 p_1 p_2 p_3^2 p_4^2 p_5+2520 p_2 p_3^3 p_4^2 p_5+2520 p_1 p_2 p_3^2 p_4 p_5^2+2520 p_2 p_3^3 p_4 p_5^2) \textcolor{red}{{z}^7}+(441 p_1^2 p_2^2 p_3^2 p_4 p_5+17172 p_1 p_2^2 p_3^3 p_4 p_5+2205 p_1 p_2^2 p_3^2 p_4^2 p_5+7875 p_1 p_2 p_3^3 p_4^2 p_5+2205 p_2^2 p_3^3 p_4^2 p_5+2205 p_1 p_2^2 p_3^2 p_4 p_5^2+7875 p_1 p_2 p_3^3 p_4 p_5^2+2205 p_2^2 p_3^3 p_4 p_5^2+1575 p_2 p_3^3 p_4^2 p_5^2) \textcolor{red}{{z}^8}+(2450 p_1^2 p_2^2 p_3^3 p_4 p_5+5600 p_1 p_2^3 p_3^3 p_4 p_5+16260 p_1 p_2^2 p_3^3 p_4^2 p_5+16260 p_1 p_2^2 p_3^3 p_4 p_5^2+5600 p_1 p_2 p_3^3 p_4^2 p_5^2+2450 p_2^2 p_3^3 p_4^2 p_5^2) \textcolor{red}{{z}^9}+(1575 p_1^2 p_2^3 p_3^3 p_4 p_5+2205 p_1^2 p_2^2 p_3^3 p_4^2 p_5+7875 p_1 p_2^3 p_3^3 p_4^2 p_5+2205 p_1 p_2^2 p_3^4 p_4^2 p_5+2205 p_1^2 p_2^2 p_3^3 p_4 p_5^2+7875 p_1 p_2^3 p_3^3 p_4 p_5^2+2205 p_1 p_2^2 p_3^4 p_4 p_5^2+17172 p_1 p_2^2 p_3^3 p_4^2 p_5^2+441 p_2^2 p_3^4 p_4^2 p_5^2) \textcolor{red}{{z}^{10}}+(2520 p_1^2 p_2^3 p_3^3 p_4^2 p_5+2520 p_1 p_2^3 p_3^4 p_4^2 p_5+2520 p_1^2 p_2^3 p_3^3 p_4 p_5^2+2520 p_1 p_2^3 p_3^4 p_4 p_5^2+3528 p_1^2 p_2^2 p_3^3 p_4^2 p_5^2+12600 p_1 p_2^3 p_3^3 p_4^2 p_5^2+5616 p_1 p_2^2 p_3^4 p_4^2 p_5^2) \textcolor{red}{{z}^{11}}+(875 p_1^2 p_2^3 p_3^4 p_4^2 p_5+875 p_1^2 p_2^3 p_3^4 p_4 p_5^2+5103 p_1^2 p_2^3 p_3^3 p_4^2 p_5^2+1225 p_1^2 p_2^2 p_3^4 p_4^2 p_5^2+9996 p_1 p_2^3 p_3^4 p_4^2 p_5^2+245 p_1 p_2^2 p_3^4 p_4^3 p_5^2+245 p_1 p_2^2 p_3^4 p_4^2 p_5^3) \textcolor{red}{{z}^{12}}+(5292 p_1^2 p_2^3 p_3^4 p_4^2 p_5^2+2016 p_1 p_2^3 p_3^5 p_4^2 p_5^2+630 p_1 p_2^3 p_3^4 p_4^3 p_5^2+630 p_1 p_2^3 p_3^4 p_4^2 p_5^3) \textcolor{red}{{z}^{13}}+(225 p_1^2 p_2^4 p_3^4 p_4^2 p_5^2+1575 p_1^2 p_2^3 p_3^5 p_4^2 p_5^2+315 p_1^2 p_2^3 p_3^4 p_4^3 p_5^2+315 p_1 p_2^3 p_3^5 p_4^3 p_5^2+315 p_1^2 p_2^3 p_3^4 p_4^2 p_5^3+315 p_1 p_2^3 p_3^5 p_4^2 p_5^3) \textcolor{red}{{z}^{14}}+(200 p_1^2 p_2^4 p_3^5 p_4^2 p_5^2+280 p_1^2 p_2^3 p_3^5 p_4^3 p_5^2+280 p_1^2 p_2^3 p_3^5 p_4^2 p_5^3+56 p_1 p_2^3 p_3^5 p_4^3 p_5^3) \textcolor{red}{{z}^{15}}+(45 p_1^2 p_2^4 p_3^5 p_4^3 p_5^2+45 p_1^2 p_2^4 p_3^5 p_4^2 p_5^3+63 p_1^2 p_2^3 p_3^5 p_4^3 p_5^3) \textcolor{red}{{z}^{16}}+18 p_1^2 p_2^4 p_3^5 p_4^3 p_5^3 \textcolor{red}{{z}^{17}}+p_1^2 p_2^4 p_3^6 p_4^3 p_5^3 \textcolor{red}{{z}^{18}}
 \)
 \item[$H_4=$]\(
 1+10 p_4 \textcolor{red}{z}+45 p_3 p_4 \textcolor{red}{{z}^2}+(70 p_2 p_3 p_4+50 p_3 p_4 p_5) \textcolor{red}{{z}^3}+(35 p_1 p_2 p_3 p_4+175 p_2 p_3 p_4 p_5) \textcolor{red}{{z}^4}+(126 p_1 p_2 p_3 p_4 p_5+126 p_2 p_3^2 p_4 p_5) \textcolor{red}{{z}^5}+(175 p_1 p_2 p_3^2 p_4 p_5+35 p_2 p_3^2 p_4^2 p_5) \textcolor{red}{{z}^6}+(50 p_1 p_2^2 p_3^2 p_4 p_5+70 p_1 p_2 p_3^2 p_4^2 p_5) \textcolor{red}{{z}^7}+45 p_1 p_2^2 p_3^2 p_4^2 p_5 \textcolor{red}{{z}^8}+10 p_1 p_2^2 p_3^3 p_4^2 p_5 \textcolor{red}{{z}^9}+p_1 p_2^2 p_3^3 p_4^2 p_5^2 \textcolor{red}{{z}^{10}}
 \)
 \item[$H_5=$]\(
 1+10 p_5 \textcolor{red}{z}+45 p_3 p_5 \textcolor{red}{{z}^2}+(70 p_2 p_3 p_5+50 p_3 p_4 p_5) \textcolor{red}{{z}^3}+(35 p_1 p_2 p_3 p_5+175 p_2 p_3 p_4 p_5) \textcolor{red}{{z}^4}+(126 p_1 p_2 p_3 p_4 p_5+126 p_2 p_3^2 p_4 p_5) \textcolor{red}{{z}^5}+(175 p_1 p_2 p_3^2 p_4 p_5+35 p_2 p_3^2 p_4 p_5^2) \textcolor{red}{{z}^6}+(50 p_1 p_2^2 p_3^2 p_4 p_5+70 p_1 p_2 p_3^2 p_4 p_5^2) \textcolor{red}{{z}^7}+45 p_1 p_2^2 p_3^2 p_4 p_5^2 \textcolor{red}{{z}^8}+10 p_1 p_2^2 p_3^3 p_4 p_5^2 \textcolor{red}{{z}^9}+p_1 p_2^2 p_3^3 p_4^2 p_5^2 \textcolor{red}{{z}^{10}}
 \)
 \end{itemize}
      
 
    \begin{center}
    {\bf Acknowledgments}
    \end{center}
  
  This paper has been supported by the RUDN University Strategic Academic Leadership Program (recipients V.D.I. and S.V.B. - mathematical and simulation model developments).  

 \end{document}